\documentclass[aps,preprint,superscriptaddress,showpacs]{revtex4}
\usepackage[english]{babel}
\usepackage{braket}
\usepackage{slashbox}
\usepackage{epsfig}
\usepackage{amsfonts}
\usepackage{graphicx}
\usepackage{hyperref}
\usepackage{natbib}
\usepackage{amsthm}
\usepackage{amsmath}
\usepackage{color}
\usepackage{times}
\usepackage{psfrag}
\usepackage{subfigure}
\usepackage{epstopdf}
\usepackage[section]{placeins}
\usepackage{multirow}
\begin{document}
\title{Probing the structural evolution along the fission path in the superheavy nucleus $^{256}$Sg}

\author{\textbf{Ting-Ting Li}}
\affiliation{School of Physics and Microelectronics, Zhengzhou
University, Zhengzhou 450001, China.}
\author{\textbf{Hua-Lei Wang}}
\email[wanghualei@zzu.edu.cn]{(Corresponding author)}
\affiliation{School of Physics
and Microelectronics, Zhengzhou University, Zhengzhou 450001, China.}
\author{\textbf{Zhen-Zhen Zhang}}
\affiliation{School of Physics and Microelectronics, Zhengzhou
University, Zhengzhou 450001, China.}
\author{\textbf{Min-Liang Liu}}
\affiliation{Key Laboratory of High Precision Nuclear Spectroscopy,
Institute of Modern Physics, Chinese Academy of Sciences, Lanzhou
730000, China.} \affiliation{School of Nuclear Science and
Technology, University of Chinese Academy of Sciences, Beijing
100049, China.}
\date{\today}

\begin{abstract}
The evolution of structure property along the fission path in the
superheavy nucleus $^{256}$Sg is predicted through the
multi-dimensional potential-energy(or Routhian)-surface calculations,
in which the phenomenological deformed Woods-Saxon potential is
adopted. Calculated nuclear deformations and fission barriers for
$^{256}_{106}$Sg$_{150}$ and its neighbors, e.g., $^{258,260}$Sg,
$^{254}$Rf and $^{252}$No are presented and compared with other
theoretical results.  A series of energy maps and curves are
provided and used to evaluate the corresponding shape-instability
properties, especially in the directions of triaxial $\gamma$ and
different hexadecapole deformations (e.g., $\alpha_{40}$,
$\alpha_{42}$ and $\alpha_{44}$). It is found that the triaxial
deformation may help the nucleus bypass the first fission-barrier of
the axial case. After the first minimum in the nuclear energy
surface, the fission pathway of the nucleus can be affected by
$\gamma$ and hexadecapole deformation degrees of freedom. In
addition, microscopic single-particle structure, pairing and
Coriolis effects are briefly investigated and discussed.

\end{abstract}
\maketitle


\section*{1. Introduction}
\label{Introduction}

The evolution of nuclear structure properties with some degree of
freedom (e.g., nucleon number, spin, temperature, etc) is one of the
most significant issues in nuclear physics~\cite{Voigt1983},
especially towards the superheavy mass region. Great progress has
been made in the synthesis of superheavy nuclei with the development
of the radioactive beam facility, heavy-ion accelerator and
highly-effective detector
systems~\cite{Liu2011,Zhao2011,Oganessian2015}. Spontaneous fission
is usually one of important decay modes in a superheavy nucleus and
the barrier along the fission path is critical to understand the
fission process~\cite{Abusara2010,Kostryukov2021}. For instance, the
survival probability of a synthesized superheavy nucleus in the
heavy-ion fusion reaction is directly related to such a barrier,
during the cooling process of a compound nucleus, which plays a
decisive role in the competition between nucleon evaporation and
fission (a small change of the fission barrier may result in several
orders of magnitude difference in survival
probability)~\cite{Lu2014}. Nevertheless, it is still rather
difficult to give an accurate description for the fission barrier so
far. To a large extent, the barrier size and shape can be determined
by the fission path in the nuclear energy surface.

Up to now, there are several types of models which are widely used
for investigating nuclear fission phenomena, including e.g., the
macroscopic-microscopic (MM) models
~\cite{Moller2009,Kowal2010,Moller2015,Gaamouci2021,Dong2015}, the
nonrelativistic energy density functionals based on zero-range
Skyrme and finite-range Gogny interactions
~\cite{Bender1998,Bonneau2004,Staszczak2009,Staszczak2007,Ling2020,Chen2022},
the extended Thomas-Fermi plus Strutinsky integral
methods~\cite{Dutta2000,Mamdouh2001} , and the covariant density
functional theory~\cite{Abusara2010,Li2010,Ring2011}. The MM methods
usually have the high descriptive power as well as simplicity of
calculation and thus are still used by many researchers so far. In
such an approach, the empirical one-body nuclear mean-filed (e.g.,
the Nilsson and Woods-Saxon potentials) Hamiltonian is used to solve
the microscopic single-particle levels and wave functions and a
macroscopic liquid-drop model (e.g., the standard liquid-drop
model~\cite{Myers1966}, the finite-range droplet
model~\cite{Moller1988}, and the Lublin-Strasboug drop
model~\cite{Pomorski2003}, etc) is combined to describe the nuclear
bulk property. In recent years, the model parameters, including
their uncertainties and propagations, in both phenomenological
Woods-Saxon potential and the macroscopic liquid-drop model are
still studied and optimized, e.g., cf
Refs.~\cite{Zhang2021,Dedes2019,Meng2022cpc,Meng2022,Gaamouci2021,Yang2022}.
Indeed, the parameters of MM models are mainly from the fitting of
available single-particle levels of several spherical nuclei and
several thousand nuclear-mass data. They are generally successful
near the $\beta$-stability line, especially in the medium and heavy
nuclear regions. Without the preconceived knowledge, e.g., about the
measured densities and single-particle energies, it may be needed to
test whether the modeling and model parameters of a phenomenological
one-body potential are still valid enough. Part of our aim of this
work is to test the theoretical method in such aspects.

Prior to this work, 16 Sg isotopes from $A=258$ to 273 were
synthesized by the fusion-evaporation reactions, e.g.,
$^{238}$U($^{30}$Si,$xn$)$^{268-x}$Sg~\cite{NNDC2022}. It was
reported that the lightest even-even Sg isotope, $^{258}$Sg, has a
revised half-life of $2.8^{+0.8}_{-0.5}$ $ms$~\cite{Heberger1997}.
Naturally, one expects that based on the fusion-evaporation
mechanism, the superheavy nuclide $^{256}$Sg will be synthesized as
the next candidate which is the nearest even-even nucleus to the
known ones in this isotopic chain. Keeping this in mind, we predict
the properties of structure evolution along the possible fission
path for the superheavy nuclide $^{256}$Sg in this project. In our
previous studies, we systematically investigated the octupole
correlation properties for 42 even-even nuclei with $102\leq Z\leq
112$~\cite{Wang2012} and the triaxial effects on the inner fission
barriers in 95 tranuranium even-even nuclei $94\leq Z\leq
118$~\cite{Chai2018}. The triaxiality and Coriolis effects on the
fission barrier in isovolumic nuclei with $A=256$ were investigated,
where the $^{256}$Sg was calculated but just focused on the first
(inner) fission barrier~\cite{Chai2018a}. In
Ref.~\cite{Chai2019CTP}, we investigated the effects of various
deformations (e.g., $\beta_2$, $\gamma$ and $\beta_4$) on the first
barrier in even-even nuclei with $N=152$ and $94 \leq Z\leq 108$. In
addition, we studied the collective rotational effects including the
$\alpha$-decay-chain nuclei (from $^{216}$Po and
$^{272}$Cn)~\cite{Chai2018IJMPE} and $^{254-258}$Rf~\cite{Wang2014}
by the similar calculation. The primary purpose of this study is to
investigate the effects of different deformation parameters,
especially the axial and non-axial hexadepole deformations, on the
fission path of $^{256}$Sg by analyzing the topography of the energy
surfaces calculated in a reasonable subspace of collective
coordinates (it is impossible to calculate in the full deformation
space). The probe of the shape evolution along the fission path on
the energy landscape will be useful for understanding the formation
mechanism of the fission barrier. We provide the analysis of the
single-particle structures, shell and pairing evolutions, especially
at the minima and saddles. Sobiczewski et al~\cite{Sobiczewski2010}
systematically investigated the static inner barrier of heaviest
nuclei with proton number $98 \leq Z \leq 126$ and neutron number
$134 \leq N \leq 192$  in a multidimensional deformation space and
pointed out that the inclusion of the non-axial hexadecapole shapes
lowers the barrier by up to about 1.5 MeV. In the synthesis of the
superheavy nuclei,  nuclear hexadecapole deformations were revealed
to have an important influence on production cross sections of
superheavy nuclei by e.g., affecting the driving potentials and the
fusion probabilities~\cite{Wang2010,Bao2016}.

This paper is organized as follows: In Sect.2, we briefly describe
the outline of the theoretical framework and the details of the
numerical calculations. The results of the calculations and their
relevant discussion are given in Sect.3. Finally, the concluding
remarks will be given in Sect.4.

\section*{2. Theoretical framework}
\label{model}

In what follows, we recall the unified procedure and give the
necessary references related to the present theoretical calculation,
which may be somewhat helpful for some readers to clarify some
details (e.g., the various variants of the pairing-energy
contribution within the framework of the macroscopic-microscopic
method). We employ potential-energy(or Routhian)-surface calculation
to study the present project. This method is based on the
macroscopic-microscopic model~\cite{Moller1995,Werner1992} and the
cranking approximation~\cite{Inglis1954,Inglis1955,Inglis1956},
which is one of widely used and powerful tools  in nuclear structure
research, especially for rotating nuclei. The usual expression for
the total energy in the rotating coordinate frame (namely, the
so-called total Routhian) reads~\cite{Nazarewicz1989}
\begin{eqnarray}
    E^{\omega}(Z,N,\hat{\beta})
    &=&
    E^{\omega}_{macr}(Z,N,\hat{\beta})
    +
    \delta E^{\omega}_{micro}(Z,N,\hat{\beta}),
                                                                  \label{eqn.01}
\end{eqnarray}
where $E^{\omega}(Z,N,\hat{\beta})$ represents the total Routhian
of a nucleus ($Z$, $N$) at frequency $\omega$ and deformation
$\hat{\beta}$. The first term on the right-hand side in
Eq.~(\ref{eqn.01}) denotes the macroscopic (liquid drop, or LD)
energy with the rigid-body moment of inertia calculated classically
at a given deformation, assuming a uniform density distribution;
$\delta E^{\omega}_{micro}$ represents the contribution due to the
microscopic effects under rotation. After rearrangement employing
elementary transformations~\cite{Bengtsson1975,Werner1995,Neergard1975,Neergard1976,Andersson1976},
the total Routhian can be rewritten as,
\begin{eqnarray}
    E^{\omega}(Z,N,\hat{\beta})
    &=&
    E^{\omega=0}(Z,N,\hat{\beta}) \nonumber \\
    &+&
    [\langle \hat{H}^{\omega}(Z,N,\hat{\beta})\rangle
    - \langle \hat{H}^{\omega=0}(Z,N,\hat{\beta})\rangle] \nonumber \\
    &-&
    \frac{1}{2}\omega^2[\mathcal{J}_{macr}(A,\hat{\beta})
    -\mathcal{J}_{Stru}(Z,N, \hat{\beta})].
                                                                  \label{eqn.02}
\end{eqnarray}
The notations for the quantities in Eq.~(\ref{eqn.02}) are
standard~\cite{Nazarewicz1989,Satula1994NPA}. The term
$E^{\omega=0}(Z,N,\hat{\beta})$ is the static total energy
(corresponding $\omega=0$) which consists of a macroscopic LD part
$E_{LD}(Z,N,\hat{\beta})$ and a shell correction $\delta
E_{shell}(Z,N,\hat{\beta})$ and a pairing-energy contribution
$\delta E_{pair}(Z,N,\hat{\beta})$ (neglecting the superscript
$\omega=0$) . The second term in the square brackets represents the
energy change of the cranked Hamiltonian
$\hat{H}^{\omega}(Z,N,\hat{\beta})$ due to
rotation~\cite{Nazarewicz1989,Satula1994NPA}. In Eq.~(\ref{eqn.02}),
it is usually and reasonably assumed that the average pairing energy
of the liquid-drop term and the Strutinsky-smeared pairing energy
cancel each other~\cite{Nazarewicz1989}. Therefore, one can further
write Eq.~(\ref{eqn.02}) as [cf. Ref.~\cite{Dudek1988} and
references therein],
\begin{eqnarray}
    E^{\omega}(Z,N,\hat{\beta})
    &=&
    E_{LD}(Z,N,\hat{\beta})                   \nonumber \\
    &+&
    \delta E_{shell}(Z,N,\hat{\beta})
    +
    \delta E_{pair}(Z,N,\hat{\beta}) \nonumber \\
    &+&
    [\langle \hat{H}^{\omega}(Z,N,\hat{\beta})\rangle- \langle
    \hat{H}^{\omega=0}(Z,N,\hat{\beta}) \rangle].
                                                                  \label{eqn.03}
\end{eqnarray}
As known, several phenomenological LD models (such as standard
liquid drop model~\cite{Myers1966}, finite-range droplet
model~\cite{Moller1995}, Lublin-Strasbourg drop
model~\cite{Pomorski2003}) with slight difference have
been developed for calculating the smoothly varying part. In these LD models,
the dominating terms are mainly associated with the volume energy, the surface energy and the Coulomb energy. In the present work,
the macroscopic energy is given by the standard LD model with
the parameters used by Myers and Swiatecki~\cite{Myers1966}.

The single-particle levels used below are calculated by solving
numerically the Schr\"{o}dinger equation with the Woods-Saxon (WS)
Hamiltonian~\cite{Dudek1980}
\begin{eqnarray}
    H_{WS}
    &=&
    T
    +
    V_{\rm cent}(\vec{r};\hat{\beta})
    +
    V_{\rm so}(\vec{r},\vec{p},\vec{s};\hat{\beta}) \nonumber \\
   && +
   V_{\rm Coul}(\vec{r},\hat{\beta}),
                                                                \label{eqn.04}
\end{eqnarray}
where the Coulomb potential $V_{\rm Coul}(\vec{r},\hat{\beta})$
defined as a classical electrostatic potential of a uniformly
charged drop is added for protons. The central part of the WS
potential is calculated as
\begin{equation}
    V_{\rm cent}(\vec{r},\hat{\beta})
    =
    \frac{V_0[1\pm\kappa(N-Z)/(N+Z)]}{1+\exp[\text{dist}_\Sigma(\vec{r},\hat{\beta})/a]},
                                                                  \label{eqn.05}
\end{equation}
where the plus and minus signs hold for protons and neutrons,
respectively and the parameter $a$ denotes the diffuseness of the
nuclear surface. The term $\text{dist}_\Sigma(\vec{r},\hat{\beta})$
represents the distance of a point $\vec{r}$ from the nuclear
surface $\Sigma$ parameterized in term of the multipole expansion of
spherical harmonics $Y_{\lambda\mu}(\theta,\phi)$ (which are
convenient to describe the geometrical properties), that is,
\begin{equation}
    \Sigma:
    R(\theta,\phi)
    =
    r_0A^{1/3}c(\hat{\beta})
    \Big[
    1
    +
    \sum_{\lambda}
    \sum_{\mu=-\lambda}^{+\lambda}
    \alpha_{\lambda\mu}
    Y^*_{\lambda\mu}(\theta,\phi)
    \Big],
                                                                  \label{eqn.06}
\end{equation}
where the function $c(\hat{\beta})$ ensures the conservation of the
nuclear volume with a change in the nuclear shape and $\hat{\beta}$
denotes the set of all the deformation parameters $\{ \alpha_{\lambda\mu} \}$. For a given nucleus with mass number $A$, a
limiting value of $\lambda<A^{1/3}$ is often estimated. In the present shape
parametrization, we consider quadrupole and hexadecapole degrees of
freedom, including nonaxial deformations, namely, $\hat{\beta}$
$\equiv$ $\{ \alpha_{20}$, $\alpha_{2\pm2}$, $\alpha_{40}$,
$\alpha_{4\pm2}$, $\alpha_{4\pm4} \}$. The quantity $R(\theta,
\phi)$ denotes the distance of any point on the nuclear surface from
the origin of the coordinate system. Because only the even $\lambda$
and even $\mu$ components are taken into account, the present
parametrisation will preserve three symmetry planes. After
requesting the hexadecpole degrees of freedom to be functions of the
scalars in the quadrupole tensor $\alpha_{2\mu}$, one can reduce the
number of independent coefficients to three, namely, $\beta_2$,
$\gamma$ and $\beta_4$, which obey the
relationships~\cite{Bhagwat2010}
\begin{equation}
  \left\{
   \begin{array}{lcl}
     \alpha_{20}
    =
     \beta_2\cos\gamma
       \\[2mm]
     \alpha_{22}
    =
    \alpha_{2-2}
    =
     -\frac{1}{\sqrt{2}}\beta_2\sin\gamma
       \\[2mm]
     \alpha_{40}
     =
     \frac{1}{6}\beta_4(5\cos^2\gamma+1)
       \\[2mm]
     \alpha_{42}
     =
     \alpha_{4-2}
     =
     -\frac{1}{12}\sqrt{30}\beta_4\sin2\gamma
       \\[2mm]
     \alpha_{44}
     =
     \alpha_{4-4}
     =
     \frac{1}{12}\sqrt{70}\beta_4\sin^2\gamma.
       \\[2mm]
   \end{array}
  \right.
                                                                  \label{eqn.07}
\end{equation}
The ($\beta_2,\gamma,\beta_4$) parametrization has all the symmetry
properties of Bohr's ($\beta_2,\gamma$)
parametrization~\cite{Bohr1952}. The spin-orbit potential,
which can strongly affects the level order, is defined by
\begin{eqnarray}
     V_{\rm so}(\vec{r},\vec{p},\vec{s};\hat{\beta})
   & = &
    -\lambda
    \Big[
    \frac{\hbar}{2mc}
    \Big]^2      \nonumber \\
    & \times & \bigg \{
    \nabla\frac{V_0[1\pm\kappa(N-Z)/(N+Z)]}{1
    +
    exp[dist_{\Sigma_{so}}(\vec{r},\hat{\beta})/a_{so}]}
    \bigg \}
    \times\vec{p}\cdot\vec{s}, \nonumber \\
                                                                  \label{eqn.08}
\end{eqnarray}
where $\lambda$ denotes the strength parameter of the effective
spin-orbit force acting on the individual nucleons. The new surface
$\Sigma_{so}$ is different from the one in Eq.~(\ref{eqn.06}) due to
the different radius parameter. In the present work, the WS
parameters are taken from Refs.~\cite{Bhagwat2010,Meng2018}, as
listed in Table~\ref{tab.wsp}.
\begin{table}[htbp]
\caption{The adopted WS parameters for both protons and neutrons
(for more details, cf e.g, Ref.~\cite{Bhagwat2010}). Note that
nuclear shape does not sensitively depend on the parameter sets in
well-deformed nuclei, especially those with large stiffness.}
\label{tab.wsp} \centering
\begin{tabular}{ccccccccccccccccccc}
\toprule V$_0$ (MeV) && $\kappa$ && r$_{0}$ (fm) && $a$(fm) & &
$\lambda$ && (r$_0$)$_{so}$ (fm) && $a_{\rm so}$ (fm)
\\
\hline
53.754 && 0.791 && 1.190&&  0.637 &&29.494 && 1.190 && 0.637  \\
\hline\hline
\end{tabular}
\end{table}


In computing the Woods-Saxon Hamiltonian matrix, the eigenfunctions
of the axially deformed harmonic oscillator potential in the
cylindrical coordinate system are adopted as the basis
functions~\cite{Cwiok1987},
\begin{equation}
|n_\rho n_z \Lambda \Sigma \rangle
  =
  \psi_{n_\rho}^\Lambda(\rho)\psi_{n_z}(z)
  \psi_{\Lambda}(\varphi)\chi(\Sigma),
                                                    \label{eqn.09}
\end{equation}
where
\begin{equation}
  \left\{
   \begin{array}{lcl}
     \psi_{n_\rho}^\Lambda(\rho)
    &=&
     \frac{\sqrt{n_{\rho}!}}{\sqrt{(n_\rho+|\Lambda|)!}}
     (2m\omega_\rho/\hbar)^{1/2} \\
     &&\times e^{-\frac{\eta^2}{2}}\eta^\Lambda
     L_{n_\rho}^{|\Lambda|}(\eta),
       \\[2mm]
     \psi_{n_z}(z)
   & =&
     \frac{1}{\sqrt{\sqrt{\pi}2^{n_z}n_z!}}
     (2m\omega_z/\hbar)^{1/4} \\
     &&\times e^{-\frac{\xi^2}{2}}
     H_{n_z}(\xi),
     \\[2mm]
     \psi_{\Lambda}(\varphi)
    & =&
     \frac{1}{\sqrt{2\pi}} e^{i\Lambda \varphi},
       \\[2mm]
   \end{array}
  \right.
                                                                  \label{eqn.10}
\end{equation}
and $\chi(\Sigma)$ represents the spin wave functions, cf. e.g.,
Sec. 3.1 in Ref.~\cite{Cwiok1987} for more details. In our
calculation, the eigenfunctions with the principal quantum number
$N \leq$ 12 and 14 have been chosen as a basis for protons and
neutrons, respectively. It is found that, by such a basis cutoff,
the results are sufficiently stable with respect to a possible
enlargement of the basis space. In addition, the time reversal
(resulting in the Kramers degeneracy) and spatial symmetries (e.g.,
the existence of three symmetry $x-y$, $y-z$ and $z-x$ planes) are
used for simplifying the Hamiltonian matrix calculation.

The shell correction $\delta E_{shell}(Z,N,\hat{\beta})$, as seen in
Eq.~(\ref{eqn.03}), is usually the most important correction to the
LD energy. Strutinsky first proposed a phenomenological expression,
\begin{equation}
\delta E_{shell}(Z,N,\hat{\beta})=\sum e_i -\int e\tilde{g}(e)de,
                                                    \label{eqn.11}
\end{equation}
where $e_i$ denotes the calculated single-particle levels and
$\tilde{g}(e)$ is the so-called smooth level density. Obviously, the
smooth level distribution function is the most important quantity,
which was early defined as,
\begin{equation}
\tilde{g}(e,\gamma)\equiv \frac{1}{\gamma\sqrt{\pi}}
                   \sum_i {\rm exp}[-\frac{(e-e_i)^2}{\gamma^2}],
                                                    \label{eqn.12}
\end{equation}
where $\gamma$ indicates the smoothing parameter without much
physical significance. To eliminate any possibly strong
$\gamma$-parameter dependence for the final result, the mathematical
form of the smooth level density $\tilde{g}(e)$ has been optimized
by introducing a phenomenological curvature-correction polynomial
$P_p(x)$\cite{Werner1995,Nilsson1969,Strutinsky1975,Ivanyuk1978}.
Then, the $\tilde{g}(e)$ expression will take the form
\begin{equation}
\tilde{g}(e,\gamma, p) = \frac{1}{\gamma\sqrt{\pi}}
                   \sum_{i=1} P_p(\frac{e-e_i}{\gamma})
                   \times{\rm exp}[-\frac{(e-e_i)^2}{\gamma^2}],
                                                    \label{eqn.13}
\end{equation}
where the corrective polynomial $P_p(x)$ can be expanded in terms of
the Hermite or Laguerre polynomials. The corresponding coefficients
of the expansion can be obtained by using the orthogonality
properties of these polynomials and Strutinsky condition (i.e., see
the APPENDIX in Ref.\cite{Pomorski2004}). In fact, this method can
be considered standard so far. For instance, the integration in
Eq.~(\ref{eqn.12}) can be calculated as follows (see
Ref.\cite{Bolsterli1972} for more details),
\begin{eqnarray}
    \int e\tilde{g}(e,\gamma,p)de
    &=&
    \int \tilde{e}(n)dn                  \nonumber \\
    &=&
    \sum_{i=1}\{\frac{1}{2}e_i [1+{\rm
    erf}(\frac{\tilde{\lambda}-e_i}{\gamma})]  \nonumber \\
    &&-
    \frac{1}{2\sqrt{\pi}}\gamma {\rm exp}[-\frac{(\tilde{\lambda}-e_i)^2}{\gamma^2}] \nonumber \\
    &&-
    \frac{1}{\sqrt{\pi}}{\rm exp}[-\frac{(\tilde{\lambda}-e_i)^2}{\gamma^2}] \nonumber \\
    &&\times \sum_{m=1}^p c_m [\frac{1}{2}\gamma H_m(\frac{\tilde{\lambda}-e_i}{\gamma})\nonumber \\
    &&+e_i H_{m-1}(\frac{\tilde{\lambda}-e_i}{\gamma}) \nonumber \\
    &&+m\gamma H_{m-2}(\frac{\tilde{\lambda}-e_i}{\gamma})]\}.
                                                                  \label{eqn.14}
\end{eqnarray}
Of course, there are some other methods developed for the shell
correction calculations, e.g., the semiclassical Wigner-Kirkwood
expansion method~\cite{Vertse1998,Bhagwat2010} and the Green's
function method~\cite{Kruppa2000}. In this work, the widely used
Strutinsky method is adopted though its known problems which appear
for mean-field potentials of finite depth as well as for nuclei
close to the proton or neutron drip lines. The smooth density is
calculated with a sixth-order Hermite polynomial and a smoothing
range $\gamma =1.20\hbar \omega _0$, where $\hbar
\omega_0=41/A^{1/3}$ MeV, indicating a satisfactory independence of
the shell correction on the parameters $\gamma$ and
$p$~\cite{Bolsterli1972}.

Besides the shell correction, the pairing-energy contribution is
also one of important single-particle corrections. Due to the
short-range interaction of nucleon pairs in time-reversed orbitals,
the total potential energy in nuclei relative to the energy without
pairing always decreases. There exist various variants of the
pairing-energy contribution in the microscopic-energy calculations,
as is recently pointed out in Ref.~\cite{Gaamouci2021}. Typically,
several kinds of the phenomenological pairing energy expressions
(namely, pairing correlation and pairing correction energies
employing or not employing the particle number projection technique)
are widely adopted in the applications of the
macroscopic-microscopic approach~\cite{Gaamouci2021}. To avoid the
confusions, it may be somewhat necessary to simply review the
`standard' definitions for pairing correlation and pairing
correction, e.g., cf Refs.~\cite{Bolsterli1972,Gaamouci2021}. For instance, the
former is given by the difference between e.g., BCS energy of the system at
pairing $\Delta \neq 0$ and its partner expression at $\Delta = 0$;
similar to the Strutinsky shell correction, the later represents the
difference between the above pairing correlation and its
Strutinsky-type smoothed out partner.

In the present work, the contribution $\delta
E_{pair}(Z,N,\hat{\beta})$ in Eq.~(\ref{eqn.03}) is the pairing
correlation energy as mentioned above. The pairing is treated by the
Lipkin-Nogami (LN) method~\cite{Pradhan1973}, which helps avoiding
not only the spurious pairing phase transition but also the particle
number fluctuation encountered in the simpler BCS calculation. In
the LN technique~\cite{Satula1994NPA,Pradhan1973}, it aims at
minimizing the expectation value of the following model Hamiltonian
\begin{equation}
    \mathcal{\hat{H}}
    =
    \hat{H}_{WS}
    +
    \hat{H}_{pair}
    -
    \lambda_1 \hat{N}
    -
    \lambda_2\hat{N}^2.
                                                                  \label{eqn.15}
\end{equation}
Here, $\hat{H}_{pair}$ indicates the pairing interaction Hamiltonian
including monopole and doubly stretched quadrupole pairing
forces~\cite{Moller1992,Xu2000,Sakamoto1990}:
\begin{equation}
    \bar{v}_{\alpha\beta\gamma\delta}^{(\lambda\mu)}
    =
   -G_{\lambda\mu}
    g_{\alpha\bar{\beta}}^{(\lambda\mu)}
    g_{\gamma\bar{\delta}}^{*(\lambda\mu)},
                                                                  \label{eqn.16}
\end{equation}
where
\begin{equation}
    g_{\alpha\bar{\beta}}^{(\lambda\mu)}
    =
    \left \{
    \begin{array}{cc}
    \delta_{\alpha\bar{}\beta} \quad  & \lambda=0, \mu=0, \\
    \langle\alpha|\widetilde{Q}_\mu |\bar{\beta}\rangle \quad  & \lambda=2,
    \mu=0,1,2.
    \end{array}
    \right.
                                                                  \label{eqn.17}
\end{equation}
The monopole pairing strength $G_{00}$ is determined by the average
gap method \cite{Moller1992} and the quadrupole pairing strengths
$G_{2\mu}$ are obtained by restoring the Galilean invariance broken
by the seniority pairing force~\cite{Sakamoto1990}. To some extent,
the quadrupole pairing can affect rotational bandhead energies,
moments of inertia, band-crossing frequencies and signature
inversion in odd-odd
nuclei~\cite{Wakai1978,Diebel1984,Satula1995,Xu2000}. The pairing
window, including dozens of single-particle levels, the respective
states (e.g. half of the particle number $Z$ or $N$) just below and
above the Fermi energy, is adopted empirically for both protons and
neutrons. The pairing gap $\Delta$, Fermi energy $\lambda$ (namely,
$\lambda_1+2\lambda_2(N_{total}+1)$), particle number fluctuation
constant $\lambda_2$, occupation probabilities $v_k^2$, and shifted
single-particle energies $\varepsilon_k$ can be determined from the
following 2$(N_2 - N_1)$ + 5 coupled nonlinear
equations~\cite{Moller1992,Pradhan1973},
\begin{equation}
  \left\{
   \begin{array}{lcl}
     N_{total}
    =
     2\sum_{k=N_1}^{N_2}v_k^2 +2(N_1-1),
       \\[2mm]
     \Delta
    =
    G\sum_{k=N_1}^{N_2}u_k v_k,
       \\[2mm]
     v_k^2
     =
     \frac{1}{2}
     \left [1-\frac{\varepsilon_k-\lambda}
     {\sqrt{(\varepsilon_k-\lambda)^2+\Delta^2}} \right ],
       \\[2mm]
     \varepsilon_k
     =
     e_k+(4\lambda-G)v_k^2,  \\[2mm]
     \lambda_2
     =
     \frac{G}{4}
    \left [
     \frac{(\sum_{k=N_1}^{N_2}u_k^3 v_k)(\sum_{k=N_1}^{N_2}u_k v_k^3)
     -\sum_{k=N_1}^{N_2}u_k^4 v_k^4}
     {(\sum_{k=N_1}^{N_2}u_k^2 v_k^2)^2
     -\sum_{k=N_1}^{N_2}u_k^4 v_k^4}
     \right ],   \\[2mm]
   \end{array}
  \right.
                                                                  \label{eqn.18}
\end{equation}
where $u_k^2=1-v_k^2$ and $k=N_1, N_1+1,\cdots, N_2$. The LN pairing
energy for the system of even-even nuclei at ``paired solution''
(pairing gap $\Delta\neq 0$) can be given
by~\cite{Pradhan1973,Moller1995}
\begin{eqnarray}
E_{LN}&=&\sum_{k}2{v_k}^2e_k-\frac{{\Delta}^2}G-G\sum_{k}{v_k}^4
\nonumber \\
     && -4{\lambda}_2\sum_{k}{u_k}^2{v_k}^2,
                                                                  \label{eqn.19}
\end{eqnarray}
where ${v_k}^2$, $e_k$, $\Delta$ and ${\lambda}_2$ represent the
occupation probabilities, single-particle energies, pairing gap and
number-fluctuation constant, respectively. Correspondingly, the
partner expression at ``no-pairing solution'' ($\Delta = 0$) reads
\begin{eqnarray}
E_{LN}(\Delta=0)&=&\sum_{k}2e_k-G\frac{N}{2}.
                                                                  \label{eqn.20}
\end{eqnarray}
The pairing correlation is defined as the difference between paired
solution $E_{LN}$ and no-pairing solution $E_{LN}$($\Delta=0$).

In the cranking calculation, we only consider the one-dimensional
approximation, supposing that the nuclear system is constrained to
rotate around a fixed axis (e.g. the $x-$axis with the largest
moment of inertia) at a given frequency $\omega$. The cranking
Hamiltonian follows the form
\begin{equation}
    H^\omega
    =
    H_{WS}
    +
    H_{pair}
    -
    \omega j_x
    -
    \lambda_1 \hat{N}
    -
    \lambda_2\hat{N}^2.
                                                                  \label{eqn.21}
\end{equation}
The resulting cranking LN equation takes the form of the well known
Hartree--Fock--Bogolyubov--like (HFB) equation which can be solved
by using the HFB cranking (HFBC) method~\cite{Ring1970} (also see,
e.g., Ref~\cite{Voigt1983}, for a detailed description). The
HFB-like equations have the following form (see, e.g.,
Ref.~\cite{Satula1994NPA}):
\begin{equation}
  \left\{
   \begin{array}{lcl}
    \sum_{\beta>0}
    \bigg \{
    \Big [
    (e_\alpha-\lambda)\delta_{\alpha\beta}
    -
    \omega (j_x)_{\alpha\beta}
    -
    G\rho^*_{\bar{\alpha}\bar{\beta}}
    +
    4\lambda_2\rho_{\alpha\beta}
    \Big ]  \\
    \times U_{\beta k}
    -
    \Delta\delta_{\alpha\beta}V_{\bar{\beta} k}
    \bigg \}
    = E_kU_{\alpha k},                                    \\ [5mm]
    \sum_{\beta>0}
    \bigg \{
    \Big [
    (e_\alpha-\lambda)\delta_{\alpha\beta}
    -
    \omega (j_x)_{\alpha\beta}
    -
    G\rho_{\alpha\beta}
    +
    4\lambda_2\rho^*_{\bar{\alpha}\bar{\beta}}
    \Big ]    \\
    \times V_{\bar{\beta} k}
    +
    \Delta^*\delta_{\alpha\beta}U_{\beta k}
    \bigg \}
    =
    E_kV_{\bar{\alpha} k},
       \end{array}
  \right.
                                                                  \label{eqn.22}
\end{equation}
where $\Delta=G\sum_{\alpha>0}\kappa_{\alpha\bar{\alpha}}$,
$\lambda=\lambda_1+ 2\lambda_2(N+1)$ and
$E_k=\varepsilon_k-\lambda_2$. Further, $\varepsilon_k$ is the
quasi-particle energy and $\alpha$ ($\bar{\alpha}$) denotes the
states of signature $r=-i$ ($r=+i$). The quantities $\rho$ and
$\kappa$ respectively correspond to the density matrix and pairing
tensor. While solving the HFBC equations, pairing is treated
self-consistently at each frequency $\omega$ and each grid point in
the selected deformation space (namely, pairing self-consistency).
Symmetries of the rotating potential are used to simplify the
cranking equations. For instance, in the present
reflection-symmetric case, both signature, $r$, and intrinsic
parity, $\pi$ are good quantum numbers. Finally, the energy in the
rotating framework can be given by
\begin{eqnarray}
  E^\omega &=& {\rm Tr}(e-\omega j_x)\rho-\frac{\Delta^2}{G}
            - G\sum_{\alpha,\beta >0}
            \rho_{\alpha,\beta}\rho_{\tilde{\alpha},\tilde{\beta}}  \nonumber \\
           && -2\lambda_2 {\rm Tr}\rho(1-\rho).
                                                              \label{eqn.23}
\end{eqnarray}
Accordingly, one can obtain the energy relative to the non-rotating
($\omega = 0$) state, as seen in the last term of
Eq.~(\ref{eqn.03}). It should certainly be mentioned that the above
derivations are used for the quasi-particle vacuum configuration of
even-even nuclear system. However, it is convenient to extend the
formalism to one or many quasi-particle excited configuration(s) by
only modifying the density matrix and pairing tensor and keeping the
form of all the equations untouched. After the numerically
calculated Routhians at any fixed $\omega$ are interpolated using,
e.g., a cubic spline function between the lattice points, the
equilibrium deformation can be determined by minimizing the
multi-dimensional potential-energy map.

\section*{3. Results And Discussion  }
\label{coherence}


\begin{figure*}[htbp]
\centering
\includegraphics[width=0.45\textwidth]{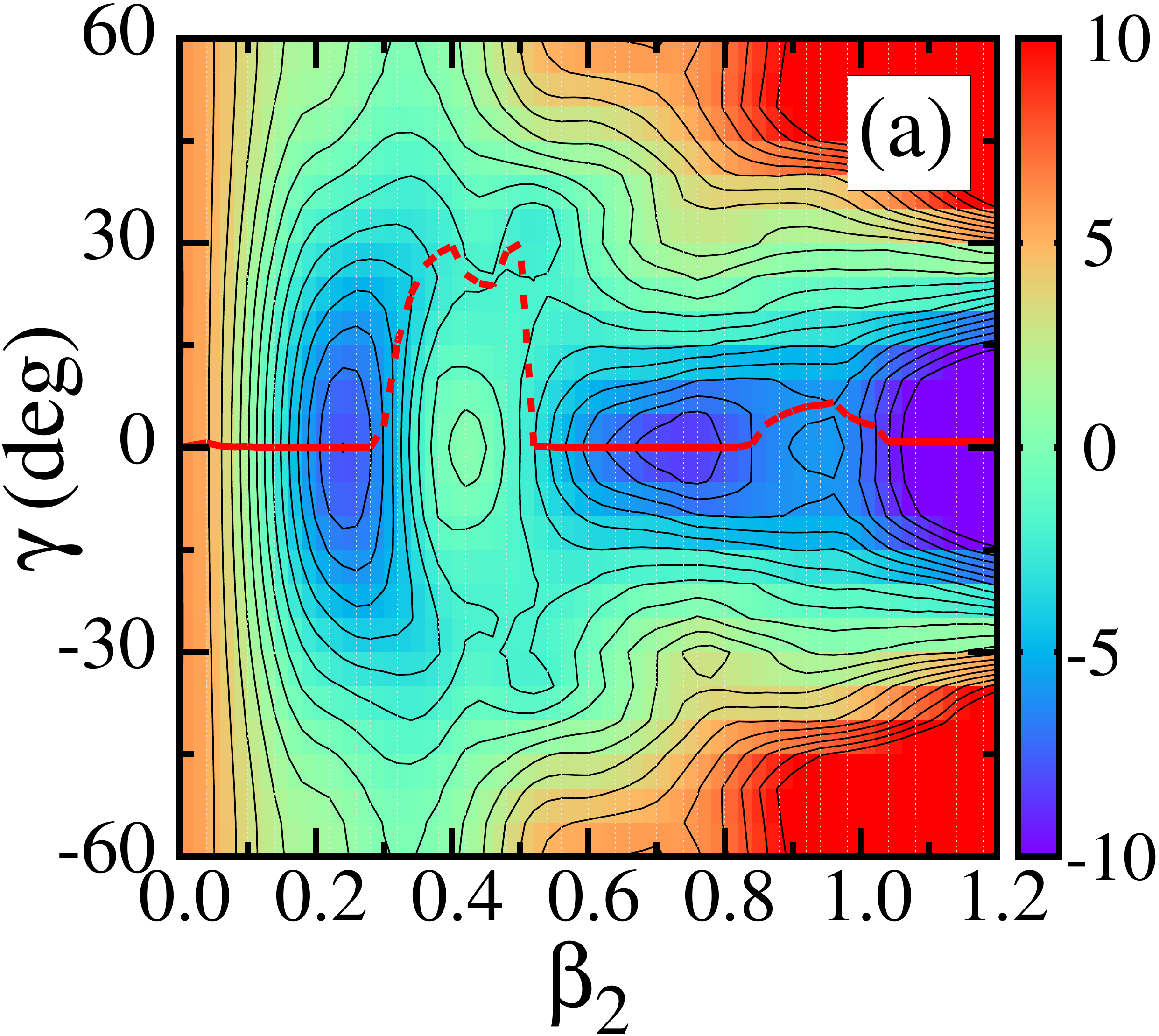}
\includegraphics[width=0.45\textwidth]{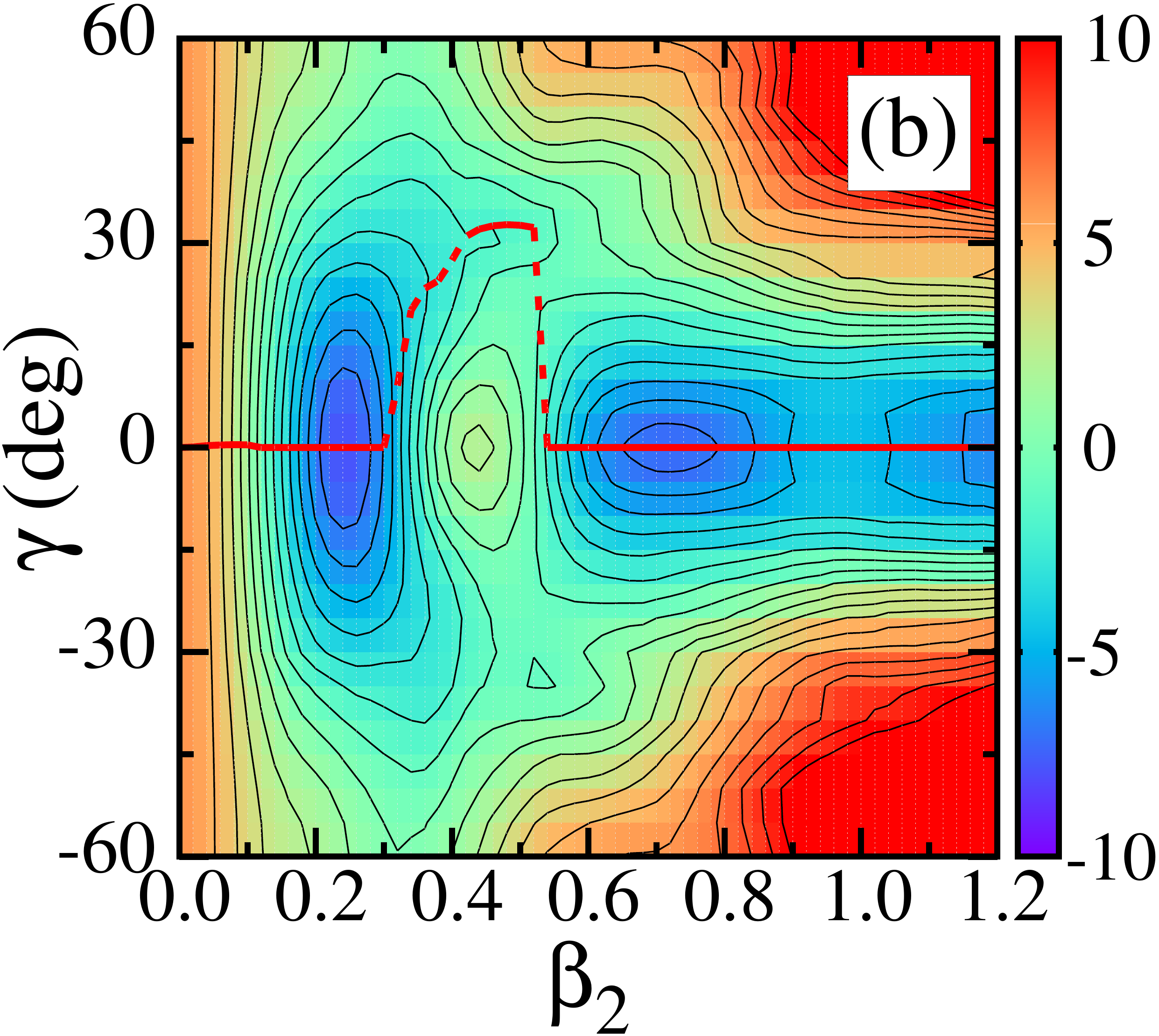}
\includegraphics[width=0.45\textwidth]{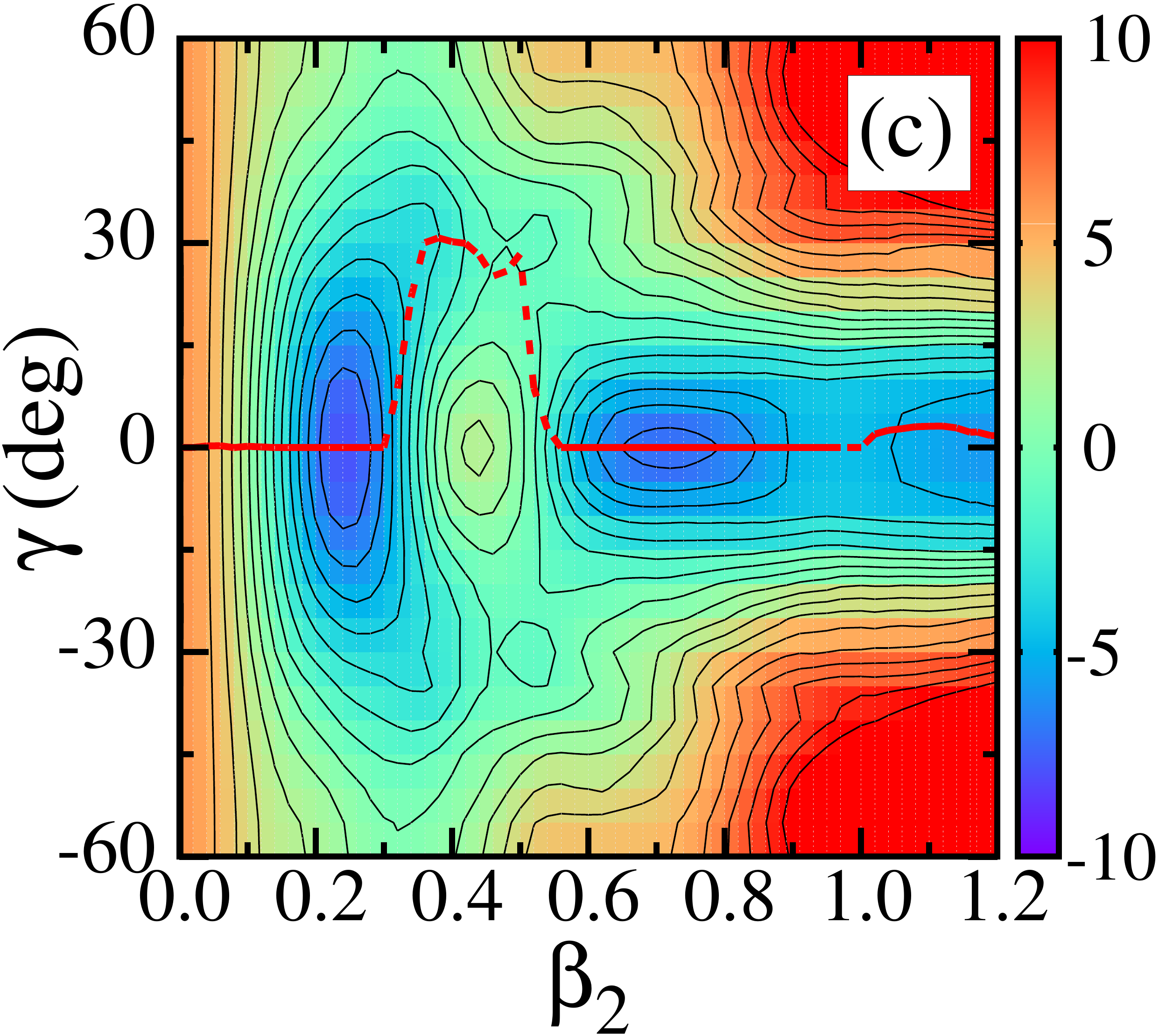}
\includegraphics[width=0.45\textwidth]{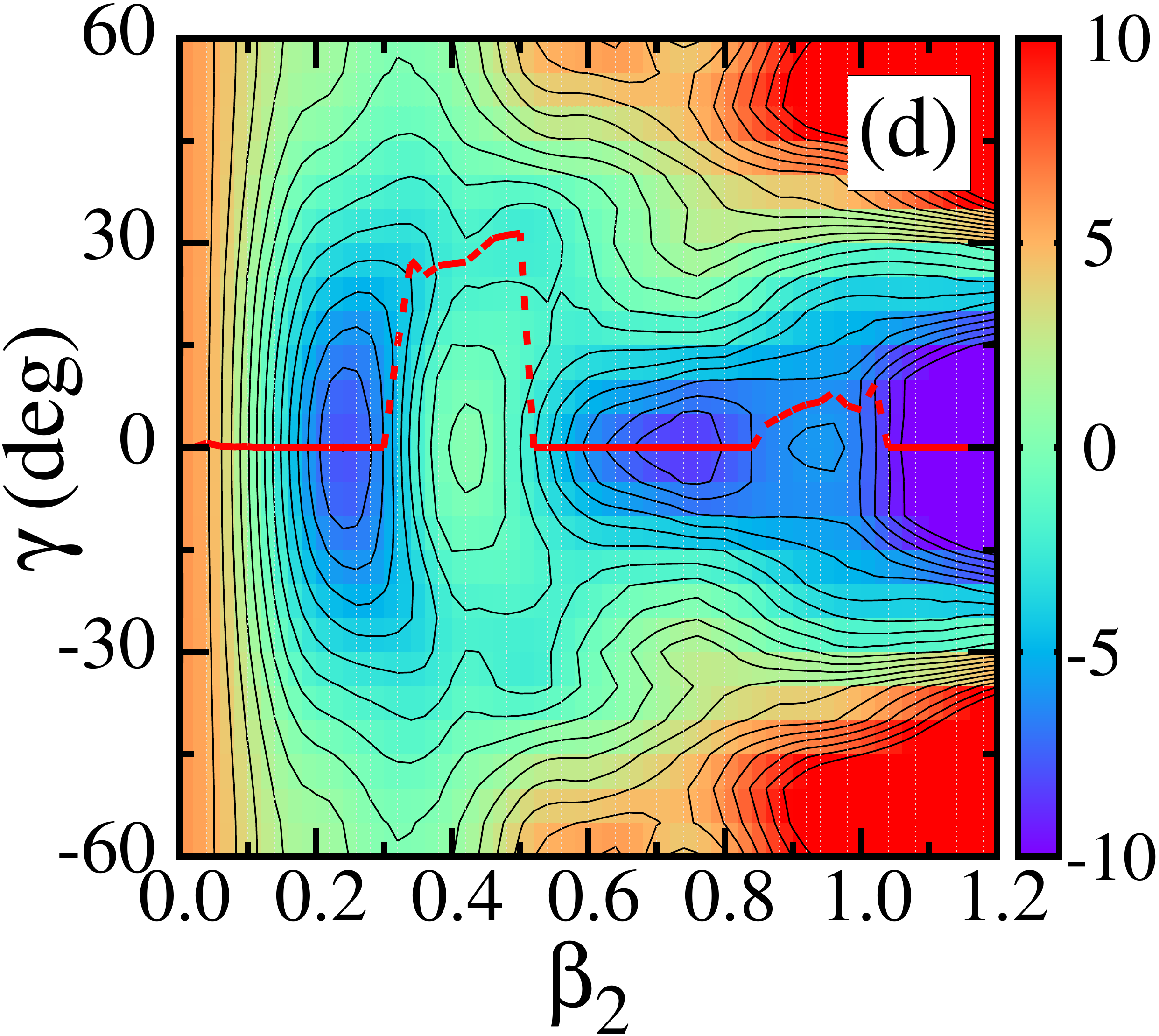}
\caption{ The projections of calculated total energy on the ($\beta_2$, $\gamma$) plane of quadrupole axial and triaxial ($\gamma$) deformations for $^{256}_{106}$Sg$_{150}$. At each deformation grid, a minimization has been performed over the hexadecapole deformation
degrees of freedom $\alpha_{40}$, $\alpha_{42}$,
$\alpha_{44}$ and $\beta_{4}$ in the subplots (a), (b), (c) and (d), respectively. The energy interval between neighbouring contour lines is 1 MeV. The red dash line denotes the possible fission pathway. See the text for more details.}
                                                     \label{Fig1}
\end{figure*}

The calculations of nuclear potential energy and/or Routhian
surfaces are very helpful for understanding the structure properties
(including the fission path) in nuclei. It is well known that
theoretical description of fission is usually based on the analysis
of the topography of the energy maps. The evolution of the potential
energy surface as a function of the collective coordinates is of
importance. We performed the nuclear potential-energy calculations
using the deformed Woos-Saxon mean-field Hamiltonian in the
deformation spaces ($\beta_2$, $\gamma$, $\alpha_{4\mu =0, 2,4}$)
and ($\beta_2$, $\gamma$, $\beta_4$). More elaborated investigation
will include the parameters related to reflection asymmetric shapes
because they are required for the description of the asymmetry in
fission-fragment mass-distribution~\cite{Zdeb2021}. In
Fig.~\ref{Fig1}, the results of potential energy surfaces projected
on ($\beta_2$, $\gamma$) plane and respectively minimized over the
hexadecapole deformation $\alpha_{40}$, $\alpha_{42}$, $\alpha_{44}$
and $\beta_{4}$ are illustrated for $^{256}_{106}$Sg$_{150}$. In
these maps, the $\beta_2$ and $\gamma$ deformation variables are
directly presented as the horizontal and vertical coordinates in a
Cartesian coordinate system, instead of the usual Cartesian
quadrupole coordinates [$X=\beta_2 \text{sin}(\gamma +30^\circ)$,
$Y=\beta_2 \text{cos}(\gamma +30^\circ)$] and the ($\beta_2$,
$\gamma$) plane in the polar coordinate system. For the static
energy surfaces, for guiding eyes, the $\gamma$ domain [$-60^\circ$,
$60^\circ$] is adopted though, in principle, half is enough. One can
see that two minima (at $\beta_2 \approx 0.24$ and 0.7) appear and
the double-humped barrier is reproduced but the second peak is lower
than those in the actinide region~\cite{Bjrnholm1980}. Calculated
energy map shows that the hexadecapole deformation has no influence
on the first minimum but can decrease the second minimum. It is
found that the $\gamma$ destroy will strongly change the fission
path, especially, between two minima.

\begin{figure*}[htbp]
\centering
\includegraphics[width=0.45\textwidth]{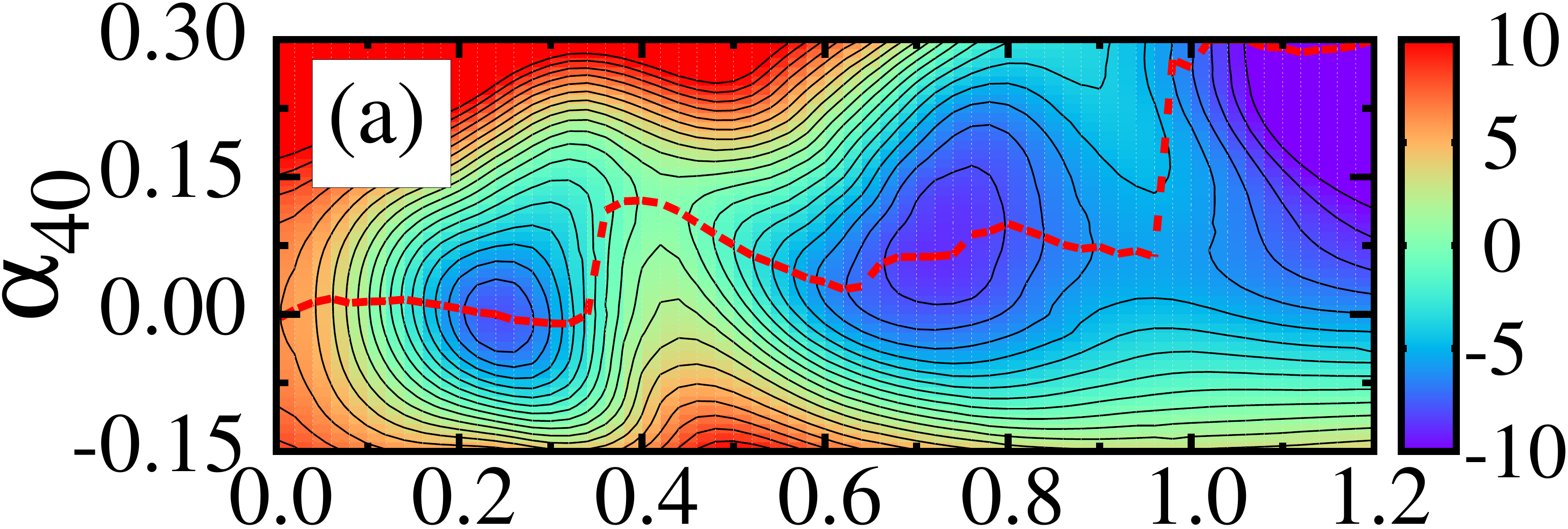}
\includegraphics[width=0.45\textwidth]{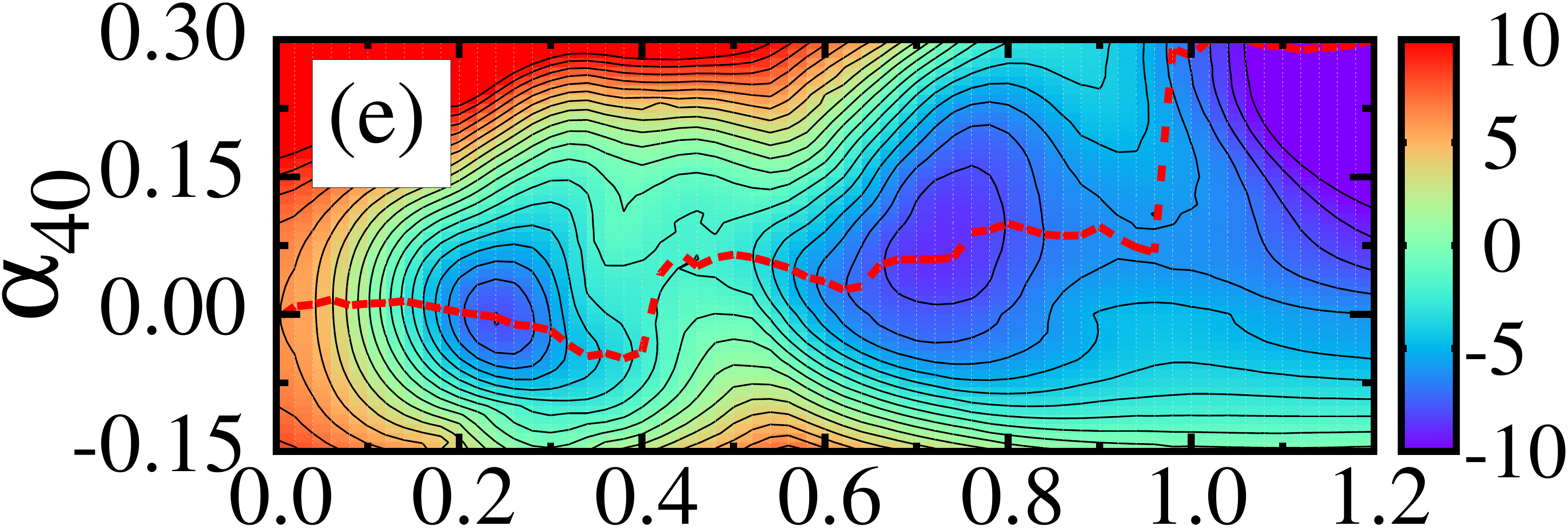}
\hspace{0.3cm}
\includegraphics[width=0.45\textwidth]{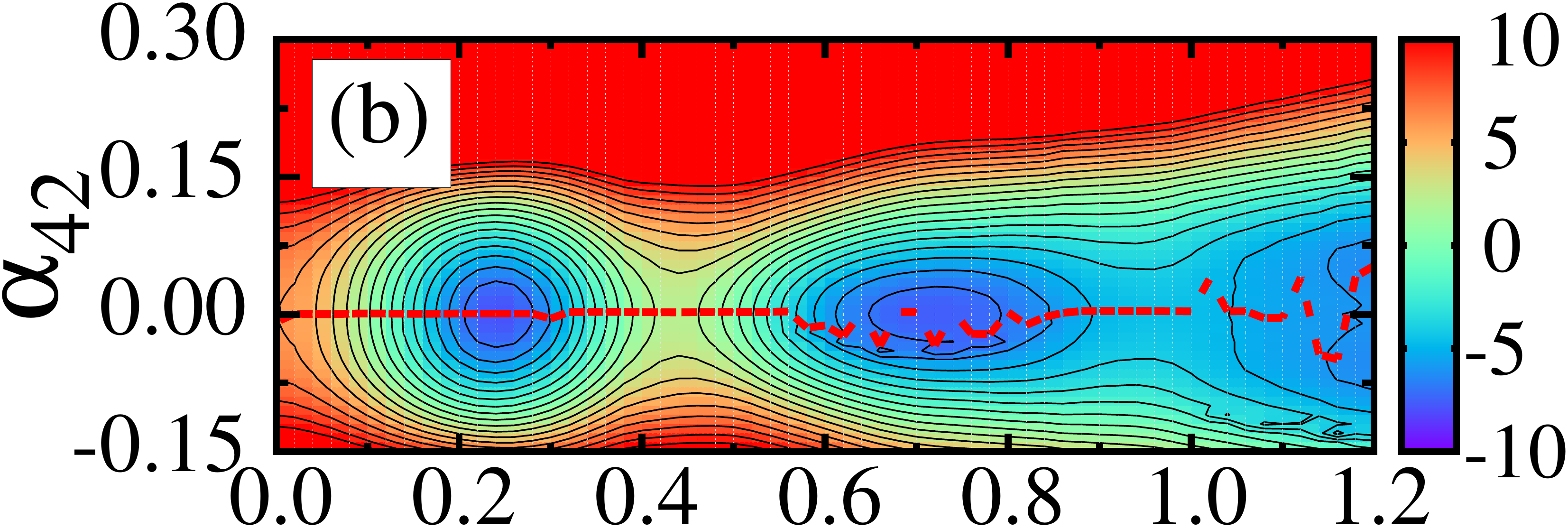}
\includegraphics[width=0.45\textwidth]{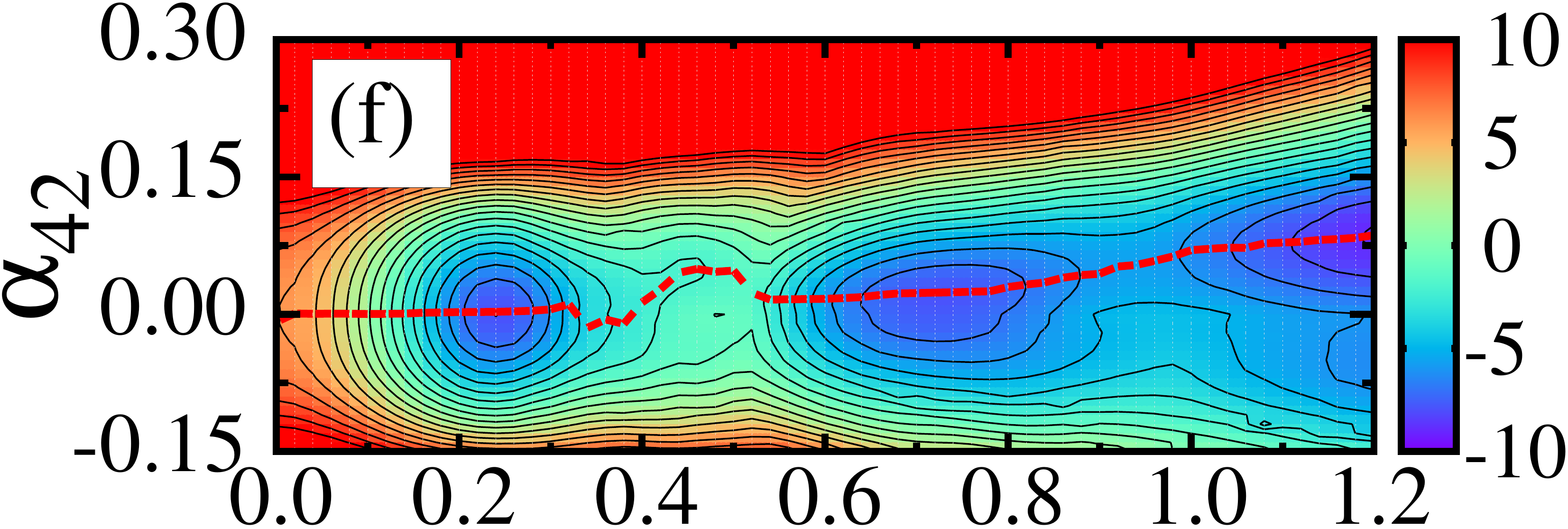}
\hspace{0.3cm}
\includegraphics[width=0.45\textwidth]{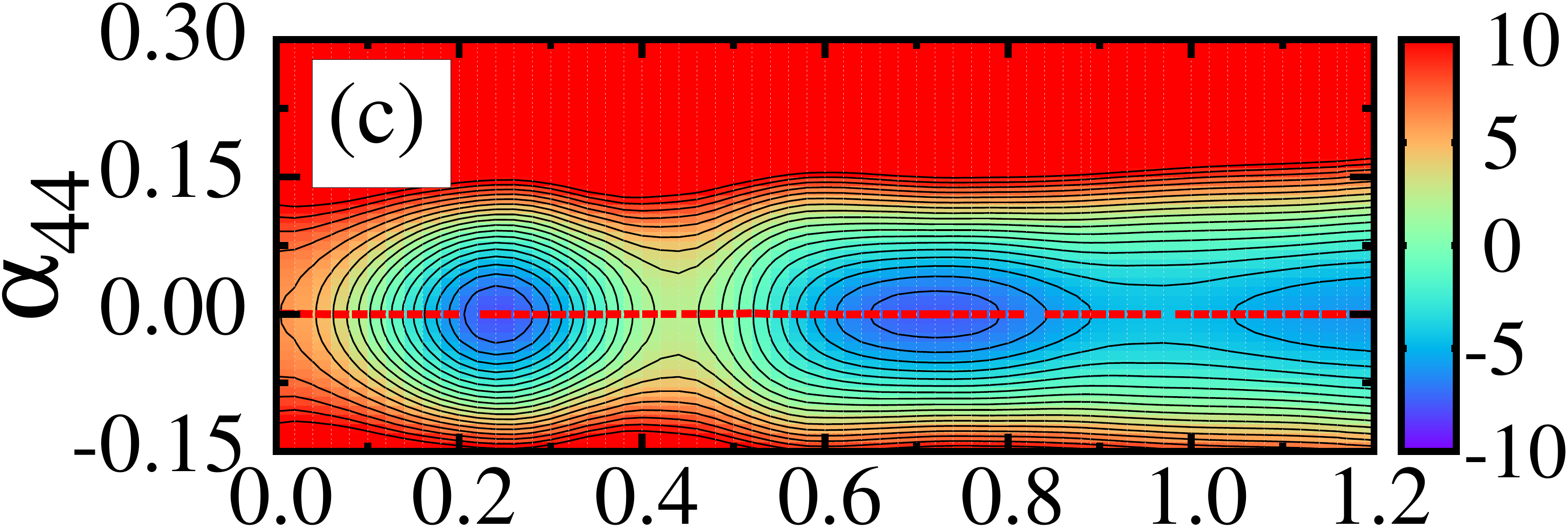}
\includegraphics[width=0.45\textwidth]{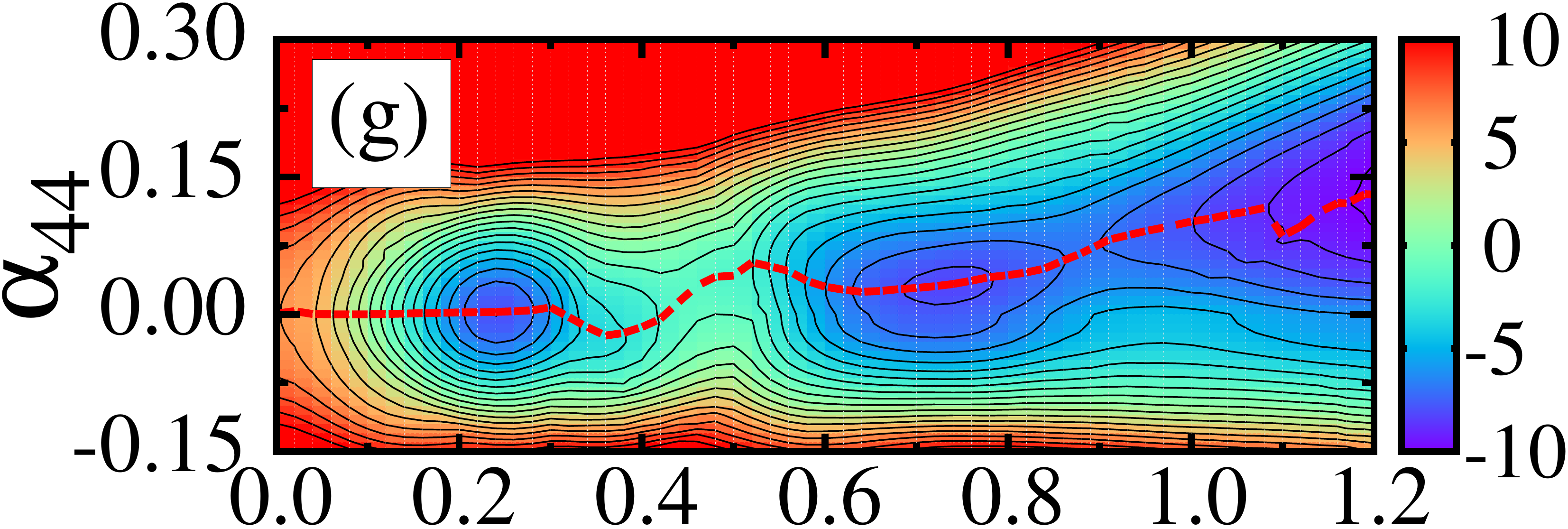}
\hspace{0.3cm}
\includegraphics[width=0.45\textwidth]{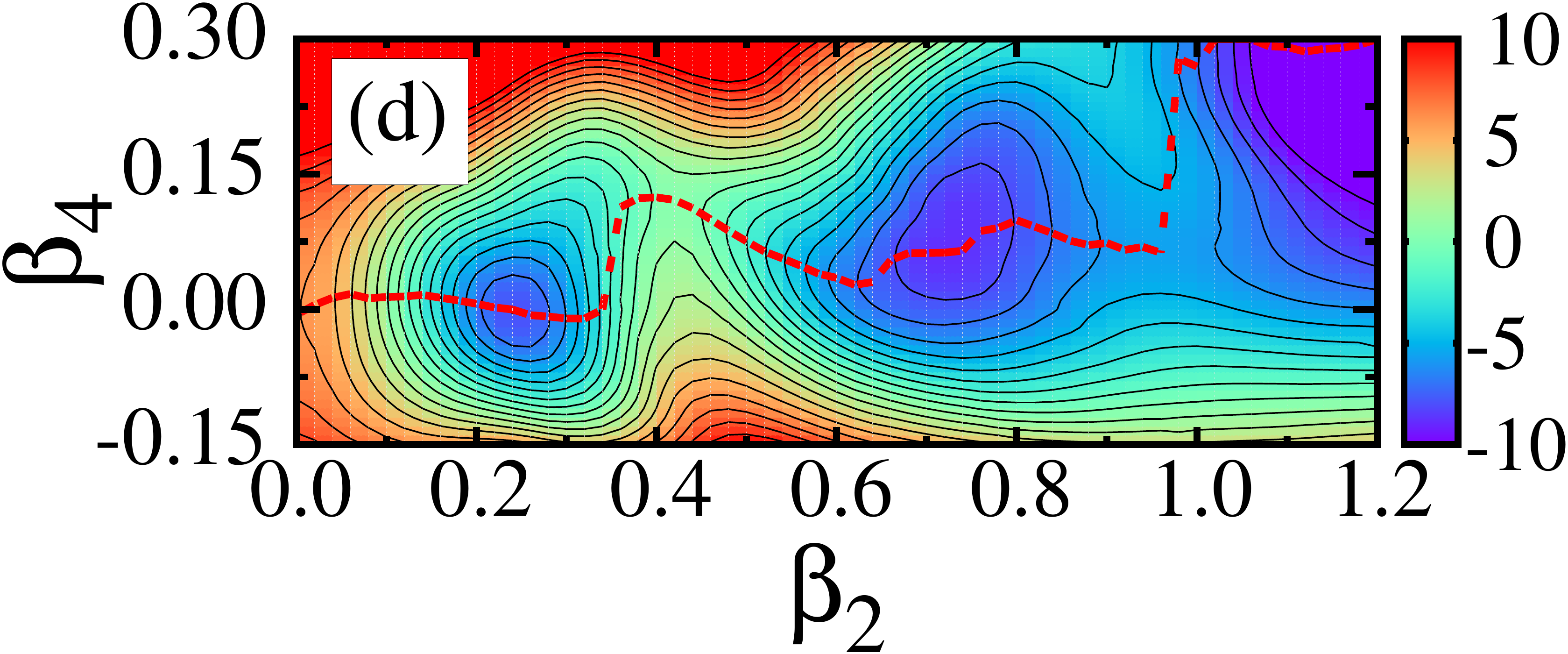}
\includegraphics[width=0.45\textwidth]{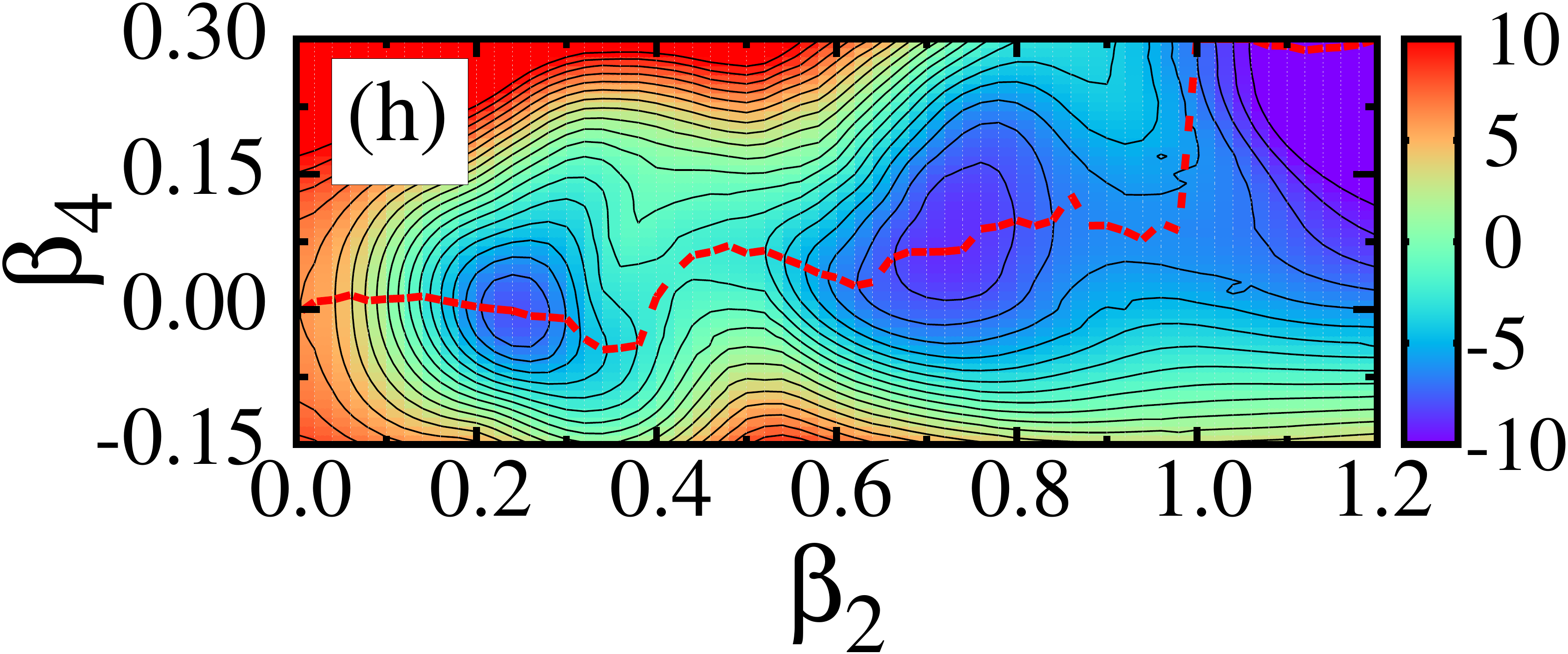}
\caption{Similar to Fig.~\ref{Fig1} but projections on ($\beta_2$,$\alpha_{40}$),
($\beta_2$,$\alpha_{42}$),($\beta_2$,$\alpha_{44}$) and
($\beta_2$,$\beta_{4}$) planes for $^{256}_{106}$Sg$_{150}$. Note that in the right four
subfigures (e),(f),(g) and (h), the minimization was performed over the triaxial deformation $\gamma$ at each mesh grid. In (a),(b),(c) and (d) subplots, the triaxial destroy was not considered. See text for more information.}
                                                     \label{Fig2}
\end{figure*}

In order to understand how dependent calculated total energies are
on these hexadecapole deformations $\alpha_{4\mu=0,2,4}$ (we focus
here on the even-$\mu$ components), Figure~\ref{Fig2} illustrates
the corresponding 2D maps projected on ($\beta_2$,
$\alpha_{4\mu=0,2,4}$) and ($\beta_2$, $\beta_4$) planes for
$^{256}_{106}$Sg$_{150}$. To separately investigate the effects of
different hexadecapole deformation parameters on the energy
surfaces, in the left four subfigures of Fig.~\ref{Fig2}, we
performed the calculations in 2D deformation spaces displayed by the
horizontal and vertical coordinates, ignoring other degrees of
freedom. It needs to be stressed that the hexadecapole deformation
$\beta_4$ involves the fixed relationships of
\{$\alpha_{4\mu=0,2,4}$\} and $\gamma$, cf. Eq.~\ref{eqn.07}. For
instance, three deformation parameters \{$\alpha_{4\mu=0,2,4}$\} can
be determined in terms of a pair of given $\beta_4$ and $\gamma$
values. It can be seen from the left panel of Fig.~\ref{Fig2} that
only $\alpha_{40}$ (equivalently $\beta_4$ at $\gamma=0^\circ$)
deformation changes the fission pathway. It seems that the non-axial
deformation parameters $\alpha_{42}$ and $\alpha_{44}$ have no
influence on the fission trajectory at this moment. In the right
part, at each deformation point of the corresponding map, the
minimization was performed over triaxial deformation $\gamma$.
Indeed, one can find that non-zero \{$\alpha_{4\mu=0,2,4}$\} values
appear along the fission pathway, indicating the three
\{$\alpha_{4\mu=0,2,4}$\} deformations play a role during the
calculations; see, e.g., Fig.~\ref{Fig2}(e)-(g). For simplicity of
calculation and simultaneously including the effects of such three
hexadecapole deformation parameters, total energy projection on the
($\beta_2$, $\beta_4$) plane is illustrated in Fig.~\ref{Fig2}(h),
minimized over $\gamma$. It was often suggested that the
3-dimensional space ($\beta_2, \gamma, \beta_4$) is the most
important, e.g., cf. Ref.~\cite{Sobiczewski2010}. Similar to the
$\gamma$ deformation, the $\beta_4$ deformation has an obvious
influence on the fission pathway after the first minimum for this
nucleus. Moreover, the $\beta_4$ deformation always keeps a non-zero
value after the first minimum.

\begin{figure*}[htbp]
\centering
\includegraphics[width=0.6\textwidth]{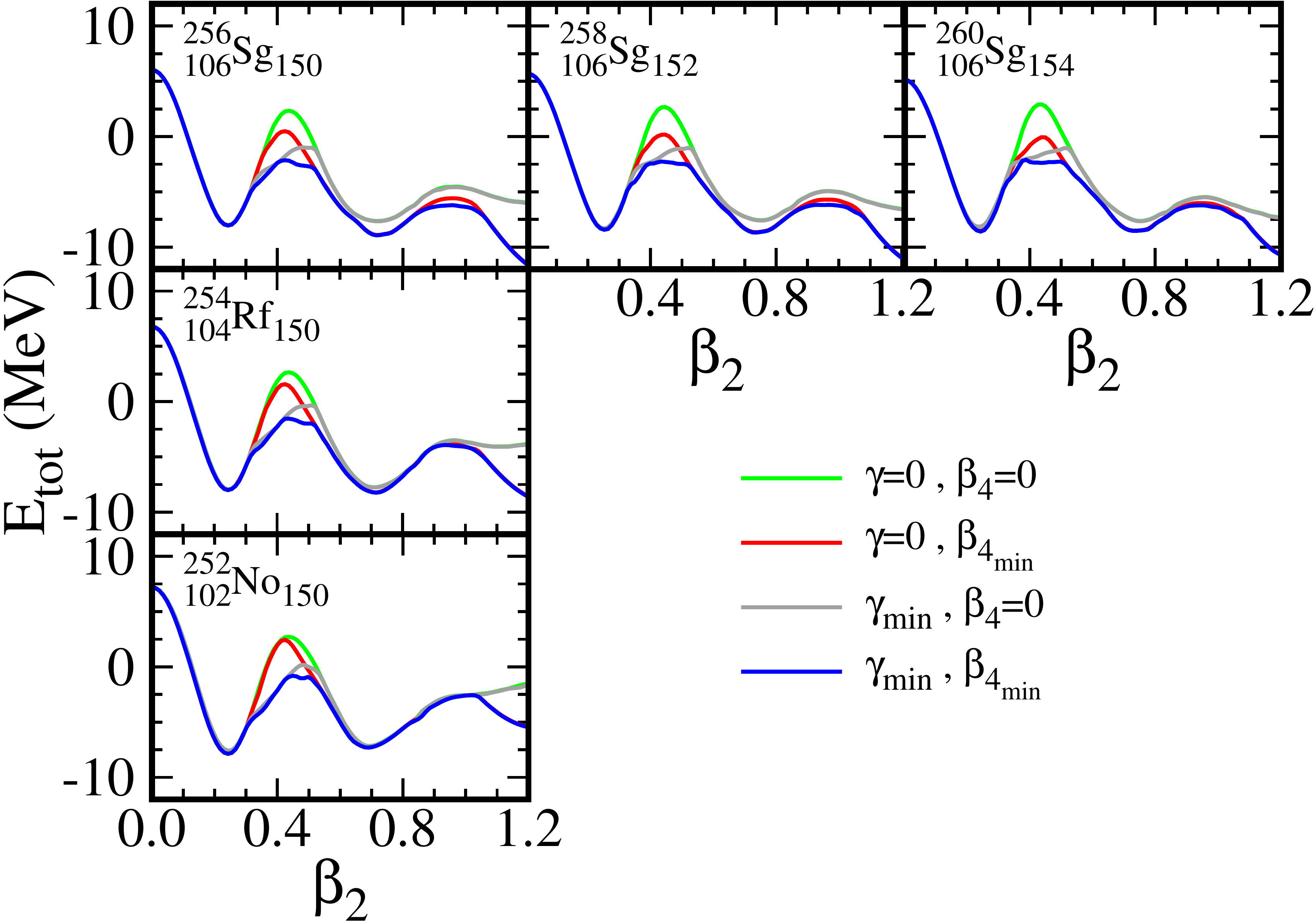}
\caption{Four types of deformation energy curves as the function of quadrupole axial deformation $\beta_2$ for
$^{256}_{106}$Sg$_{150}$ and its two isotopic
and isotonic neighbours, namely, $^{258}_{106}$Sg$_{152}$,
$^{260}_{106}$Sg$_{154}$, $^{254}_{104}$Rf$_{150}$ and
$^{252}_{102}$No$_{150}$. At each $\beta_2$ point, the minimization was performed over $\gamma$ and/or $\beta_4$. The legends denote that whether or not  total energy at each $\beta_2$ was minimized and, if so, with respect to what deformation parameter(s). See text for further explanations.}
                                                     \label{Fig3}
\end{figure*}

From the 2D energy $\beta_2$ vs $\gamma$ and $\beta_2$ vs $\beta_4$
maps, we can obtain the further energy projection e.g., on the
$\beta_2$ direction. By such an operation, the total energy curve
will be given, which is usually useful for extracting the
information of fission barrier. Figure~\ref{Fig3} illustrates four
types of total energy curves in functions of $\beta_2$ for five
selected nuclei $^{256,258,260}$Sg, $^{254}$Rf and $^{252}$No. Note
that the blue, grey, red and green lines respectively correspond to
those curves whose energies are minimized over $\gamma$ and
$\beta_4$; $\gamma$; $\beta_4$; and none. By them, one can see the
evolution of the energy curves from both isotopic and isotonic
directions. It seems that from the isotonic direction,
$^{256}_{106}$Sg$_{150}$ is the critical nucleus in which the
hexadecapole deformation $\beta_4$ always play a role after the
first minimum. From this figure, we can obtain the equilibrium
deformations of different minima and maxima, further the height of
fission barriers. The impact of the triaxial and hexadecapole
deformations on the energy curves can clearly evaluated.  The
inclusion of different deformation parameters can affect not only
the height but also the shape of the fission barrier. As noted in
Ref.~\cite{Zdeb2021}, the tunneling probability through the fission
barrier will depend exponentially on the square root of its height
times its width, when approximated by a square potential barrier.
One can find that the triaxial deformation can decrease the barrier
hight, especially for the inner barrier e.g. in $^{256}$Sg.
Nevertheless, the hexadecapole deformation (responsible for
necking~\cite{Tsekhanovich2019}) decreases both the height and the
width of the fission barrier. Even, as seen in $^{256,258}$Sg, the
least-energy fission path is strongly modified by the hexadecapole
deformation after their first minima. After the second saddles, the
effect of the hexadecapole deformation becomes significant in all
selected nuclei. However, it was found that the octupole deformation
will play an important role at the second saddle and after that,
leading to a change of the obtained mass asymmetry at the scission
point~\cite{Wang2012,Zdeb2021,Lu2014}.

\begin{table*}
\caption{\label{tab:table1}The results of potential-energy-surface
(PES) calculations for ground-state equilibrium deformation
parameter $\beta_2$ and inner fission barriers $B_f$ for the 5
selected even-even nuclei, together with some other theoretical
calculations for comparison; see the text for more descriptions.}
\begin{ruledtabular}
\begin{tabular}{cccccccccc}
\multirow{3}{*}{Nuclei}&\multicolumn{5}{c}{$\beta_2$}&\multicolumn{4}{c}{$B_f$/MeV} \\
\cline{2-6}   \cline{7-10}
 &PES&HN~\cite{Sobiczewski2001}&FF~\cite{Moller2016}&HFBCS~\cite{Goriely2001}&ETFSI~\cite{Aboussir1995}& PES& HN~\cite{Kowal2010} &FFL~\cite{Moller2009}&ETFSI~\cite{Mamdouh2001} \\
$^{260}_{106}$Sg$_{154}$&0.243&0.247&0.242&0.31&0.25&6.49&6.28&5.84&4.6\\
$^{258}_{106}$Sg$_{152}$&0.242&0.247&0.252&0.27&0.25&6.16&6.22&5.93&4.7\\
$^{256}_{106}$Sg$_{150}$&0.243&0.246&0.252&0.25&0.27&5.88&5.46&5.30&---\\
$^{254}_{104}$Rf$_{150}$&0.243&0.247&0.252&0.27&0.27&6.44&5.74&5.87&5.3\\
$^{252}_{102}$No$_{150}$&0.243&0.249&0.250&0.30&0.26&7.01&6.52&6.50&5.8\\
\end{tabular}
\end{ruledtabular}
\end{table*}

In Table~\ref{tab:table1}, the present results (calculated
quadrupole deformation $\beta_2$ and fission barrier $B_f$) for five
selected nuclei are confronted with other accepted theories (the
experimental data are scarce so far), including the results of the
heavy-nuclei (HN) model~\cite{Sobiczewski2001,Kowal2010}, the
fold-Yukawa (FY) single-particle potential and the finite-range
droplet model (FRDM)~\cite{Moller2016}, the Hartree-Fock-BCS
(HFBCS)~\cite{Goriely2001}, the fold-Yukawa (FY) single-particle
potential and the finite-range liquid-drop model
(FRLDM)~\cite{Moller2009}, and the extended Thomas-Fermi plus
Strutinsky integral (ETFSI)~\cite{Aboussir1995,Mamdouh2001} methods.
Comparison shows that these results are somewhat model-dependent but
in good agreement with each other to a large extent. It can be found
that the HFBCS calculation gave the larger equilibrium deformations
and our calculation has the higher inner fission-barriers. Our
calculated deformations may be underestimated to some extent, cf.
Ref.~\cite{Zhang2022}. As discussed by Dudek et
al.~\cite{Dudek1984}, the underestimated quadrupole deformation
$\beta_2$ should be slightly modified by the empirical relationship
$1.10\beta_2$-$0.03(\beta_2)^3$. Within the framework of the same
model, it can be seen that the selected five nuclei almost have the
same $\beta_2$ in the PES, HN and FF (FY+FRDM)~\cite{Moller2016}
calculations. In the HFBCS and ETFSI calculations, the nucleus
$^{256}$Sg has the largest and the smallest $\beta_2$ values in the
five nuclei, respectively, but the differences are still rather
small. Concerning the inner fission barriers, it seems that the
present calculation may relatively overestimate the barriers.
However, the present calculation has the same trends to the results
given by HN and FFL (FY+FRLDM)~\cite{Moller2009} calculations. For
instance, the nucleus $^{256}$Sg has the smallest inner barrier in
these five nuclei, in good agreement with those in HN and FFL
calculations. In our previous publication~\cite{Chai2018}, a lower
$B_f$ about 4.8 MeV was obtained by using the universal parameter
set. This value is lower about 1 MeV than the present calculation
(5.88 MeV, as seen in the table) and lower than the values by HN and
FFL calculations. The further experimental information is desirable.
Interestingly, though the inner barrier of $^{256}$Sg is the lowest,
its outer barrier ($\sim 2.72$ MeV) is higher than those in its
isotopic neighbors $^{258,260}$Sg ($\sim 2.52$ and 2.29 MeV). It is
certainly expected that the outer barrier of $^{256}$Sg can
relatively increase the survival probability of this superheavy
nucleus, benifiting for the observation in experiment to some
extent.

\begin{figure}[htbp]

\centering
\includegraphics[width=0.45\textwidth]{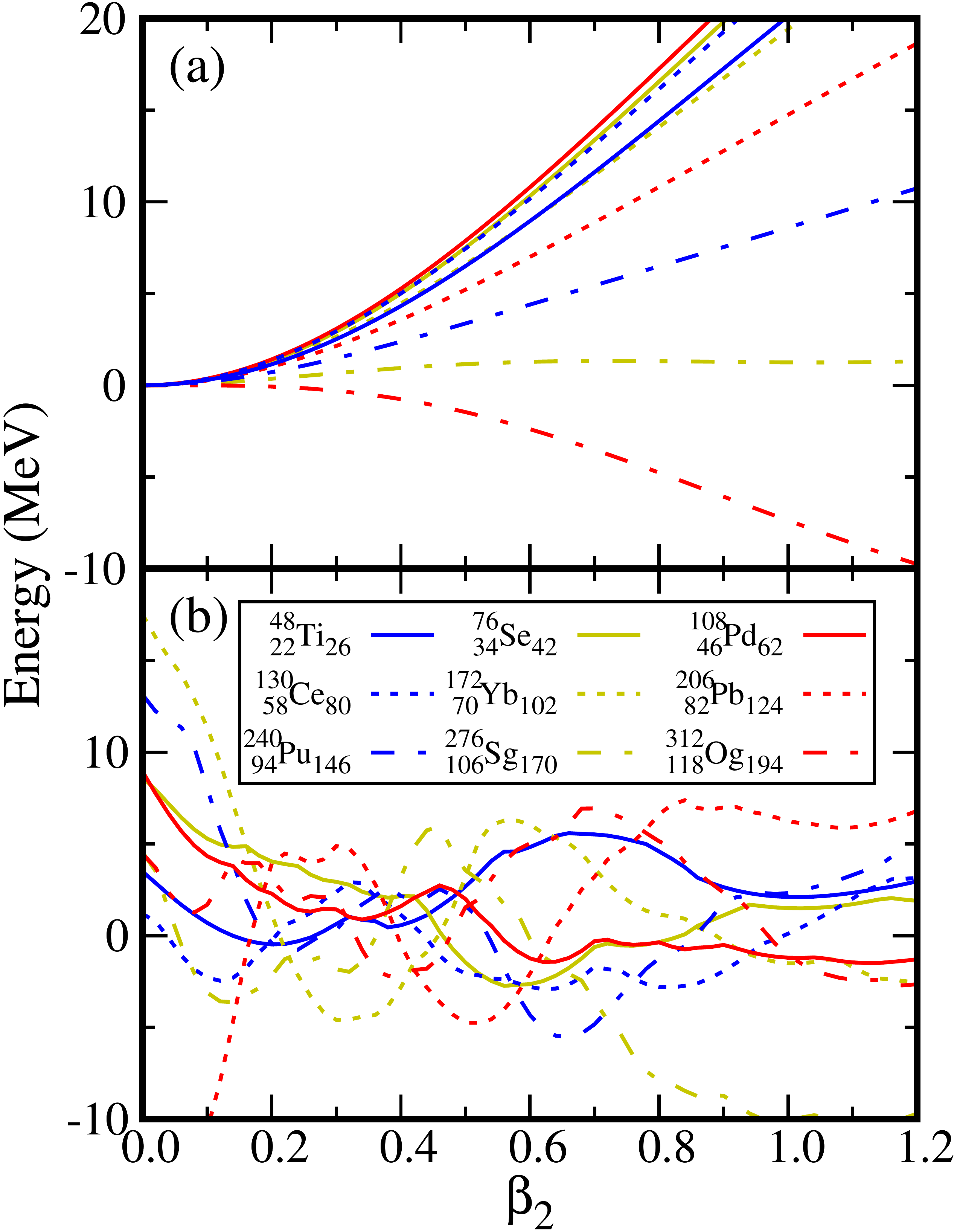}
\caption{Macroscopic energies (a) and Shell correction energies (b) as the function of quadrupole axial deformation $\beta_2$ for
several selected nuclei (see the legends, or cf. Ref.~\cite{Meng2022}) along the $\beta$-stability line. Note that during the calculation other deformation parameters are set to be zero.}\vspace{-0.5cm}
                                                     \label{Fig4}
\end{figure}

In macroscopic-microscopic model, as is well known, the total energy
is mainly determined by the liquid-drop energy and shell correction.
In Fig.~\ref{Fig4}, to understand their evolution properties from
light to heavy nuclei, we show the macroscopic energy and
microscopic shell correction for arbitrarily selected nine nuclei
along the $\beta$-stability line (cf. Ref.~\cite{Meng2022}). As
excepted, one can see that with increasing mass number $A$ the
macroscopic energy (the important contribution of fission barrier)
is decreasing at a given $\beta_2$ (e.g., $\sim 0.4$, about the
position of the first barrier;cf. Fig.~\ref{Fig3}) deformation,
indeed, almost approaching zero in the superheavy region [e.g., with
$Z \gtrsim 104$, see $^{276}_{106}$Sg$_{170}$ in Fig.~\ref{Fig4}(a),
indicating the disappearance of the macroscopic fission barrier]. In
particular, the calculated liquid-drop energy rapidly descends with
increasing $\beta_2$ in the ``heavier'' superheavy nucleus
$^{312}_{118}$Sg$_{194}$ which denotes that it is more difficult to
bound such a heavy nucleon-system. Figure~\ref{Fig3}(b) illustrates
the corresponding shell corrections for the selected nuclei
mentioned above. Indeed, the energy staggering is rather large and
combining the smoothed macroscopic energy, the potential pocket(s)
can appear, which is the formation mechanism of superheavy nuclei.

\begin{figure}[htbp]
\centering
\includegraphics[width=0.45\textwidth]{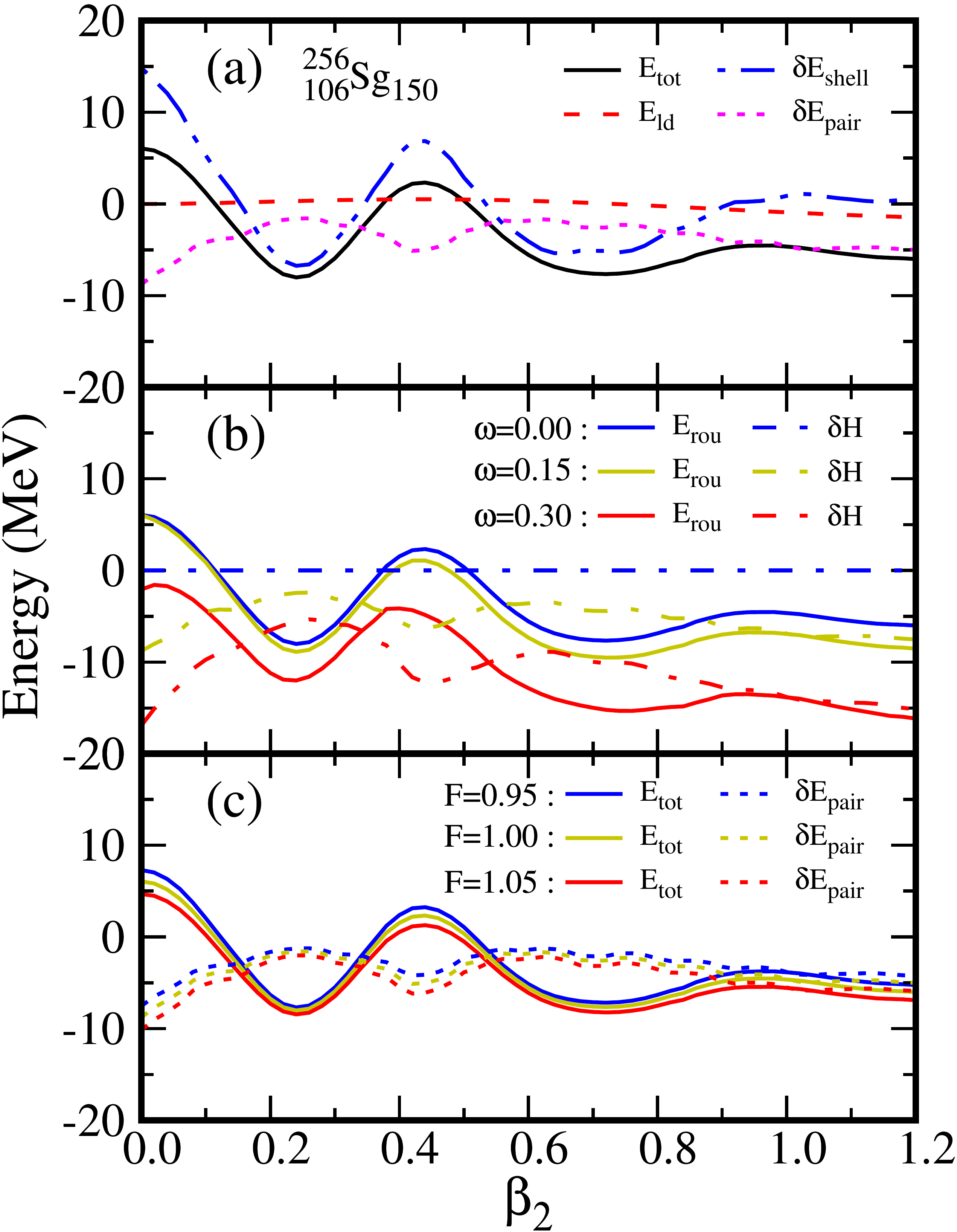}
\caption{(a) Total energy $E_{tot}$ curve (together
with its macroscopic liquid-drop energy $E_{ld}$ and microscopic shell correction and pairing correlation energies, namely, $\delta E_{shell}$ and $\delta E_{pair}$) vs $\beta_2$ deformation for the nucleus
$^{256}_{106}$Sg$_{150}$. For simplicity, other deformation degrees of freedom were closed during the calculation. (b) Similar to (a) but for the total Routhian ($E_{rou}$) curves  and the corresponding rotational contribution $\delta $H at three selected frequencies $\hbar\omega=0.00, 0.15$ and 0.30 MeV. (c)Similar to (a) but for the total energy and the corresponding pairing correlation $\delta E$ at three selected pairing-strength factor $F=0.95, 1.00$ and 1.05 (the adjusted pairing strength $G=FG_0$).}
                                                     \label{Fig5}
\end{figure}

\begin{figure}[htbp]
\centering
\includegraphics[width=0.48\textwidth]{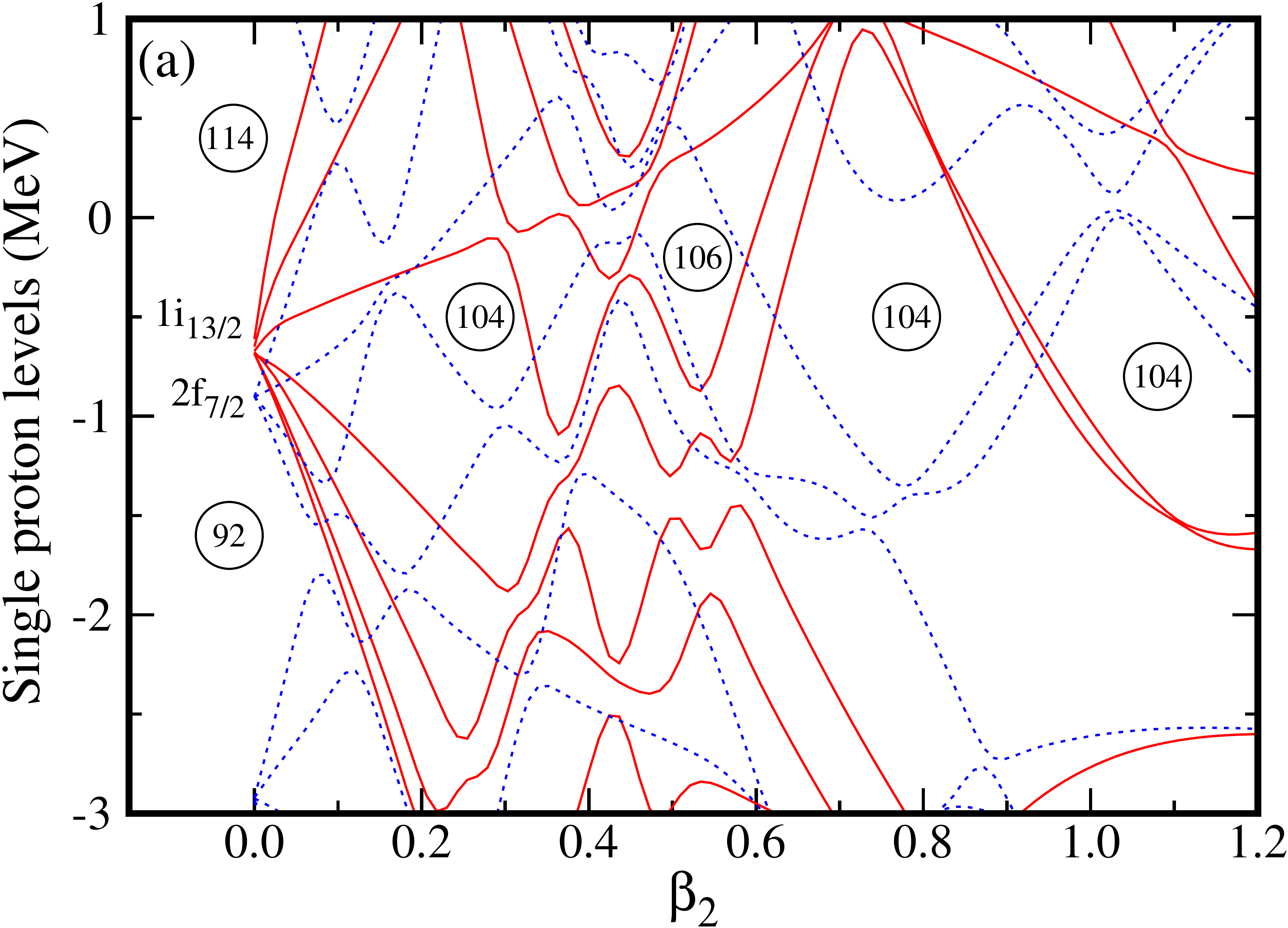}
\includegraphics[width=0.48\textwidth]{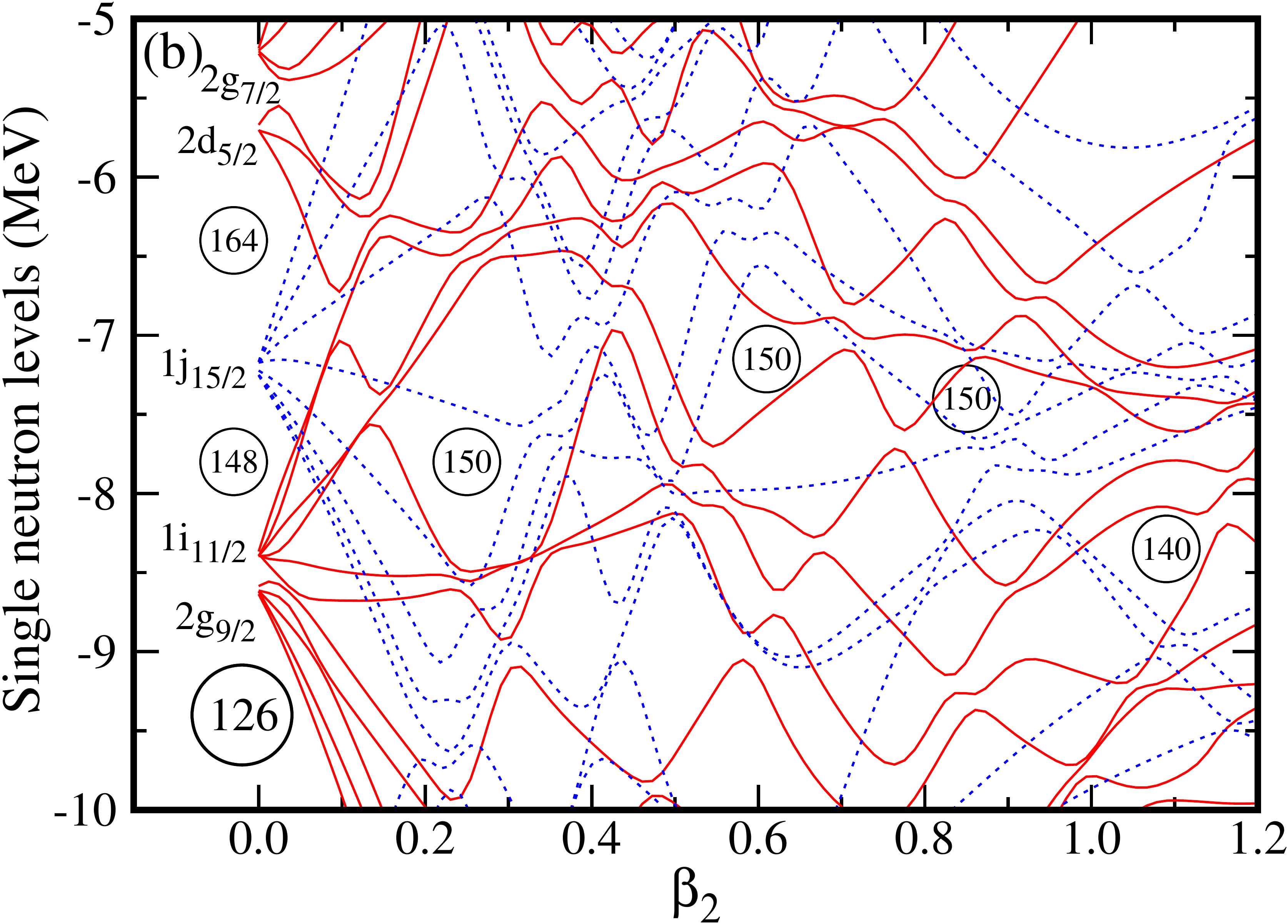}
\caption{Calculated proton (a) and neutron (b) single-particle energies as functions of the quadrupole deformation $\beta_2$ for $^{256}_{106}$Sg$_{150}$, focusing on the domain near the Fermi surface. The levels with positive and negative parities are respectively denoted by red solid and blue dotted lines. Spherical single-particle orbitals (i.e., at $\beta_2$ = 0.0)
in the window of interest are labeled by the quantum numbers $nlj$.}
                                                     \label{Fig6}
\end{figure}

In Fig.~\ref{Fig5}, we provide the further evolution information on the total energy and its different components in functions of the quadrupole deformation $\beta_2$ for $^{256}_{106}$Sg$_{150}$. Figure~\ref{Fig5}(a) illustrates that total energy, together with the macroscopic liquid-drop energy $E_{ld}$, shell correction $\delta E_{shell}$ and pairing correlation $\delta E_{pair}$. For simplicity, other deformation degrees of freedom are ignored. In this nucleus, as seen, the macroscopic energy fully makes no contribution to the fission barrier. The barrier is mainly formed by the quantum shell effect. The inclusion of short-range pairing interaction always decreases the total energy, showing an irregular but relatively smoothed change (decreasing the barrier here). With increasing $\beta_2$, the shell effect tends to disapear. In the subfigure Fig.~\ref{Fig5}(b), we show the total Routhian and the rotational contribution at ground-state and two selected frquencies $\hbar\omega=$ 0.15 and 0.30 MeV, aiming to see the effect of the Coriolis force. One can see that, similar to the trend of the pairing correlation, the energy due to rotation will decrease the barrier because the energy difference e.g., at the positions of the first barrier and the first minimum is a negative value. It should be noted that the selected rotational frequencies respectively correspond to the values before and after the first band-crossing frequency in such a normal-deformed superheavy nucleus, e.g., cf. Ref.~\cite{Wang2014}. Along the curve, the ground-state or yrast configuration for the nucleus may be rather different (see, e.g., Fig.~\ref{Fig6}, the occupied single-particle levels below the Fermi surface will generally be rather different). In Fig.~\ref{Fig5}(c), the total energy and its pairing correlation energy are illustrated with different pairing strengths by adjusting the factor $F$ (e.g., in $G=FG_0$, where $G_0$ is the orginal pairing strength). It can be noticed that the pairing correlation energy will decrease with increasing pairing strength $G$. Both at the barrier and the minimum, the effects seem to be very similar. At the large deformation region, the pairing correlation tends to a constant.


The microscopic structure of  nuclei is primarily determined by the
single-particle levels, especially near the Fermi
level~\cite{Baldo2020}. Experimentally, one can detect and
investigate single-particle states by e.g., the inelastic electron
scattering [like $(e, e^\prime p)$], the direct stripping and
pick-up reactions [typically $(p, d)$ and $(d, p)$ reactions],
$\beta$-decay rates, and so on~\cite{Bertsch1983,Vaquero2020}.
Because the measured single-particle states may be not pure, a
rigorous definition of these states is given by the Green's function
formalism (cf. Ref.~\cite{Baldo2020}), showing that it is necessary
to extract the spectroscopic factor. Such a quantity will provide an
illustration of how much a single-particle level can be considered
as a pure state and whether or not the correlations (e.g., the
short- and long-range ones) beyond the mean field appear.
Theoretically, the single-particle levels correspond to the
eigenstates of the mean-field Hamiltonian (e.g., the
Woods-Saxon-type one in this work). They are also the building
blocks of the many-body wave functions, e.g., in self-consistent
Hartree-Fock calculation. In Fig.~\ref{Fig6}, the single-particle
levels near the proton and neutron Fermi surfaces are respectively
illustrated in (a) and (b) parts. A set of conserved quantum numbers
(associated with a complete set of commuting observables) are
usually used for labeling the corresponding single-particle levels
and wave functions. For instance, the spherical single-particle
levels are denoted by the spherical quantum numbers $n, l$ and $j$
(corresponding the principal quantum number, the orbital angular
momentum, and the total angular momentum, respectively). Similar to
atomic spectroscopy, the notations $s$, $p$, $d$, $f$, $g$, $h$
$\cdots$ (corresponding to $l = $ $0$, $1$, $2$, $3$, $4$, $5$
$\cdots$, respectively) are used. Due to the strong spin-orbit
coupling, the single particle state with $l$ will split into two
states with $j$ = $l$ $\pm$ 1/2 (The degeneracy of each spherical
single-particle level can be calculated by $2j+1$). In the present
work, one can see that the expected shell structure and shell
closure can be well reproduced. When deformed shape occurs, the
$2j$+1 degeneracy will be broken and the spherical single-particle
level will split into $j+1/2$ components (each one is typically
double degenerate due to Kramers degeneracy). These deformed
single-particle levels are generally described by asymptotic Nilsson
quantum numbers $\Omega^\pi [Nn_z\Lambda]$, where $N$ is the total
oscillator shell quantum number; $n_z$ stands the number of
oscillator quanta in the $z$ direction (the direction of the
symmetry axis); $\Lambda$ is the projection of angular momentum
along the symmetry axis; $\Sigma$ is the projection of intrinsic
spin along the symmetry axis; $\Omega$ is the projection of total
angular momentum $j$ (including orbital $l$ and spin $s$) on the
symmetry axis and $\Omega=\Lambda+\Sigma$. Note that the Nilsson
labels are not given owing to space limitations. Similar to magnetic
field, in the rotational coordinate system, the Coriolis force
resulted from the non-inertial reference frame can also break the
time reversal symmetry and mix the Nilsson states. Then, the
single-particle Routhians can only be labeled by the conserved
parity and signature $(\pi, \alpha)$ or $(\pi, r)$ (cf.
Ref.~\cite{Voigt1983} for the rigorous definition). It should be
pointed out here that we did not perform the virtual crossing
removal~\cite{Bengtsson1985} of single-particle levels with same
symmetries in these plots but this will not affect the
identification of the single-particle levels. From Fig.~\ref{Fig6},
one can see that the shell gaps appear at the energy-minimum
positions with lower level-densities and the higher level-densities
occur at the saddle positions (cf. e.g., Fig.~\ref{Fig5}). The
deformed neutron shells at $N=152$ and 162 are
reproduced~\cite{Oganessian2015}.

\begin{figure*}[htbp]
\centering
\includegraphics[width=0.45\textwidth]{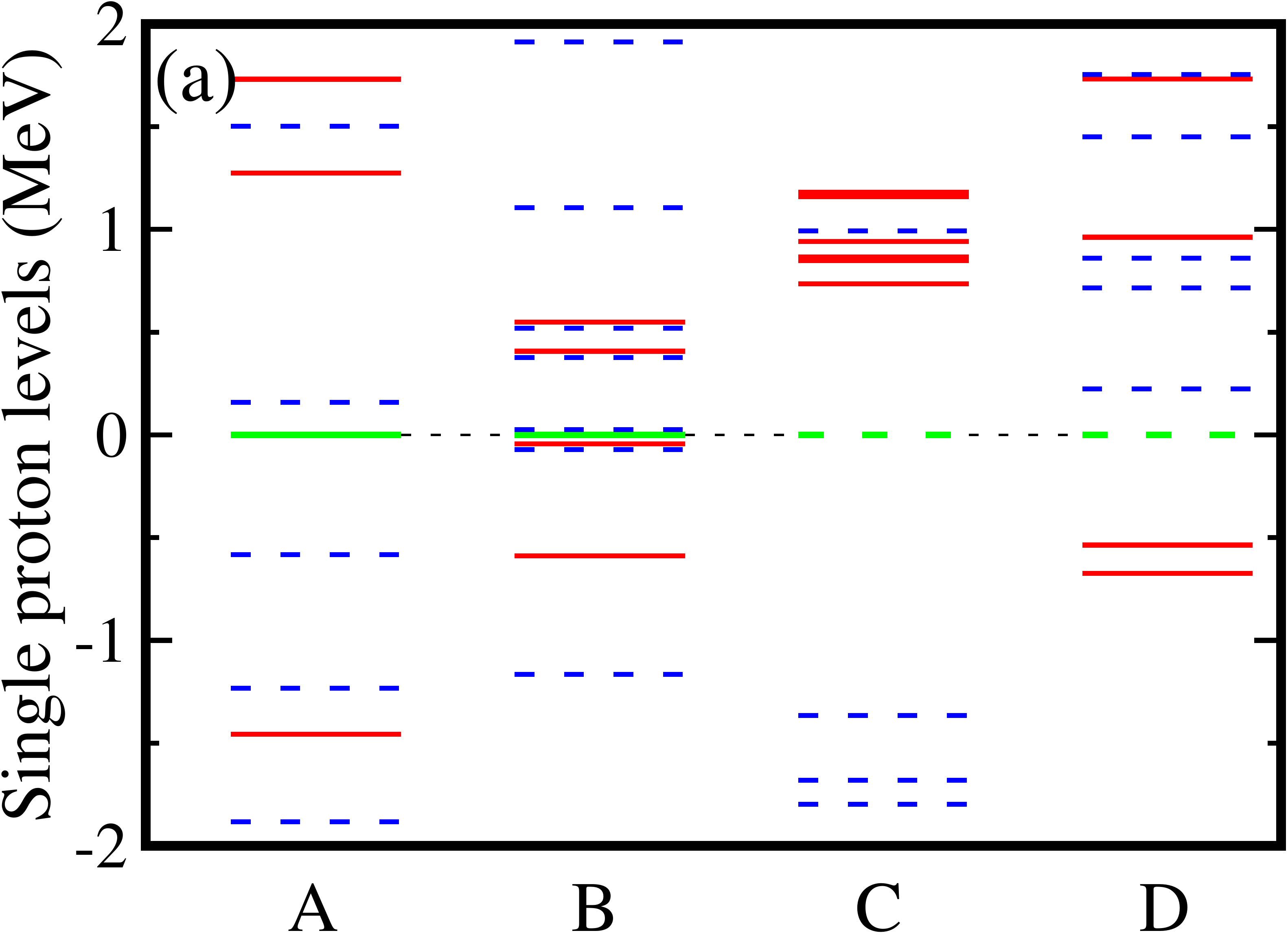}
\includegraphics[width=0.45\textwidth]{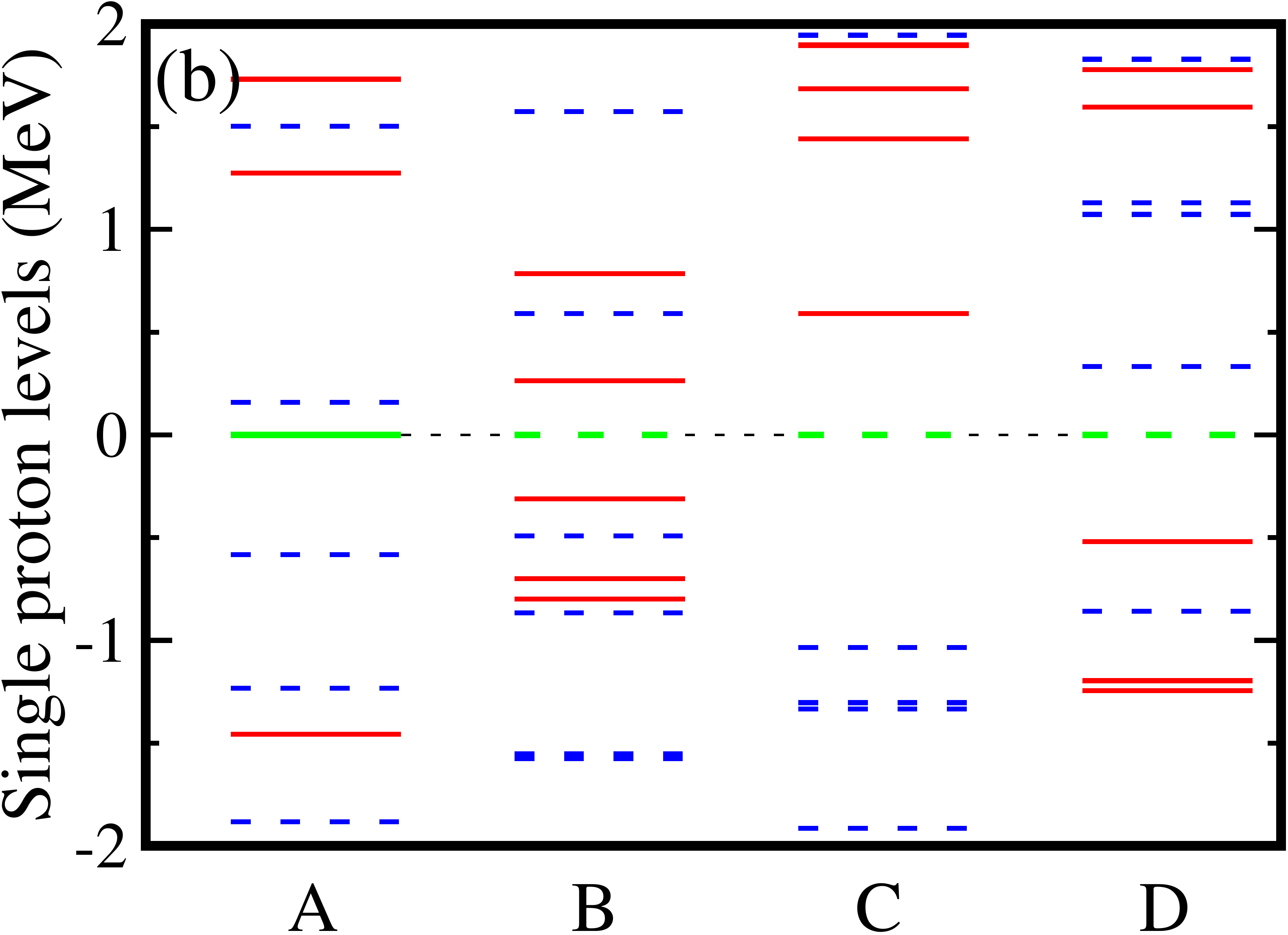}
\includegraphics[width=0.45\textwidth]{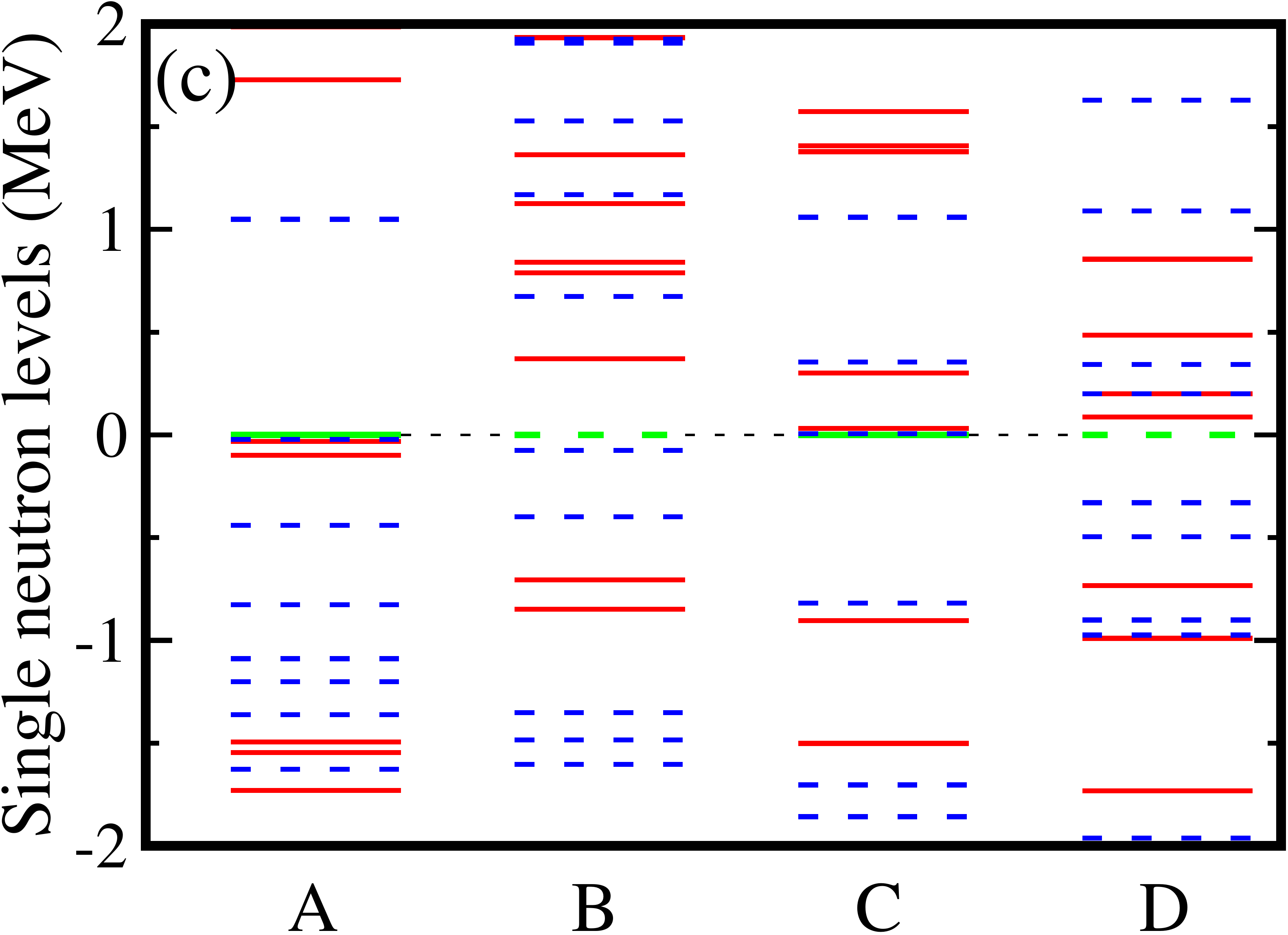}
\includegraphics[width=0.45\textwidth]{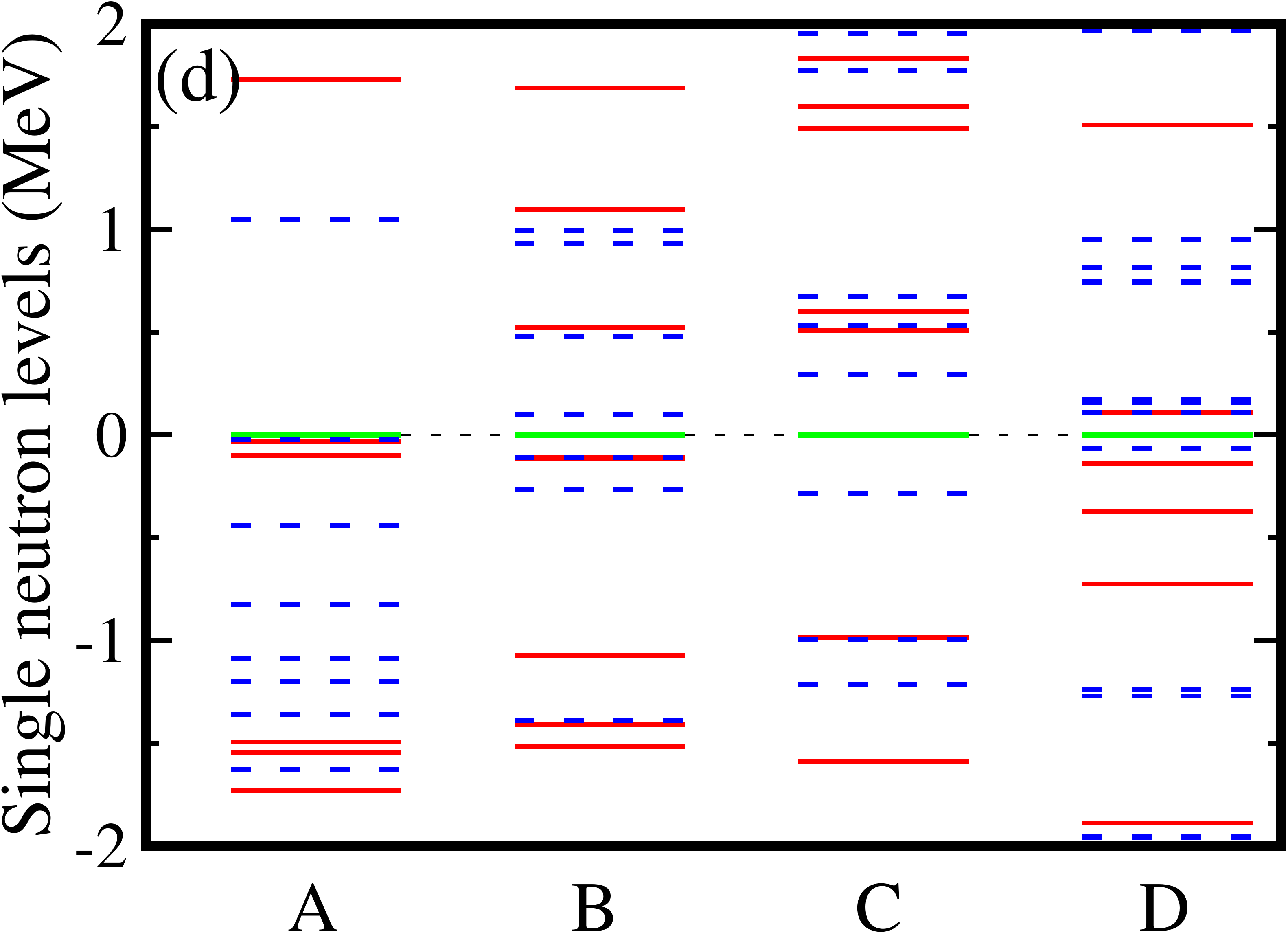}
\caption{(a) Calculated proton single-particle levels for
$^{256}_{106}$Sg$_{150}$ at the four typical $\beta_2$ deformation points ($A$, the 1st minimum; $B$, the 1st maximum;, $C$, the 2nd minimum; and $D$, the 2nd maximum) along the energy curve , see e.g., Fig.~\ref{Fig3}. In this plot, only $\beta_2$ deformation is considered for simplicity, corresponding to the blue energy curve in Fig.~\ref{Fig3}. (b) Similar to (a) but, in this plot, the energy is minimized over $\beta_4$ for each $\beta_2$ points, corresponding to the red energy curve in Fig.~\ref{Fig3}. (c) Similar to (a) but for neutron single-particle levels. (d) Similar to (b) but for neutron single-particle levels.}
                                                     \label{Fig7}
\end{figure*}

For a clear display about the level density near the minimum and
saddle points, Figure~\ref{Fig7} presents the proton and neutron
single-particle levels at these corresponding deformation points.
Note that the Fermi levels (the green levels) at the four typical
points $A, B, C$ and $D$ are shifted to zero for comparison. The
levels in Fig.~\ref{Fig7}(a) and (c) correspond to deformation
conditions same to those in Figs.~\ref{Fig5} and \ref{Fig6} where
only the $\beta_2$ deformation is considered. In the right two
subfigures of Fig.~\ref{Fig7}, at each $\beta_2$ point, the
``realistic'' $\beta_4$ value is taken into account (the equilibrium
deformation is adopted after potential-energy minimization over
$\beta_4$). Relative to the left two ones, the levels are rearranged
to an extent by the hexadecapole deformation degree of freedom. As
excepted, the level density is lower (higher) near the minimum
(saddle) point, indicating the occurrence of a largely negative
(positive) shell correction.
%
%
%

\begin{figure*}[htbp]
\centering
\includegraphics[width=0.45\textwidth]{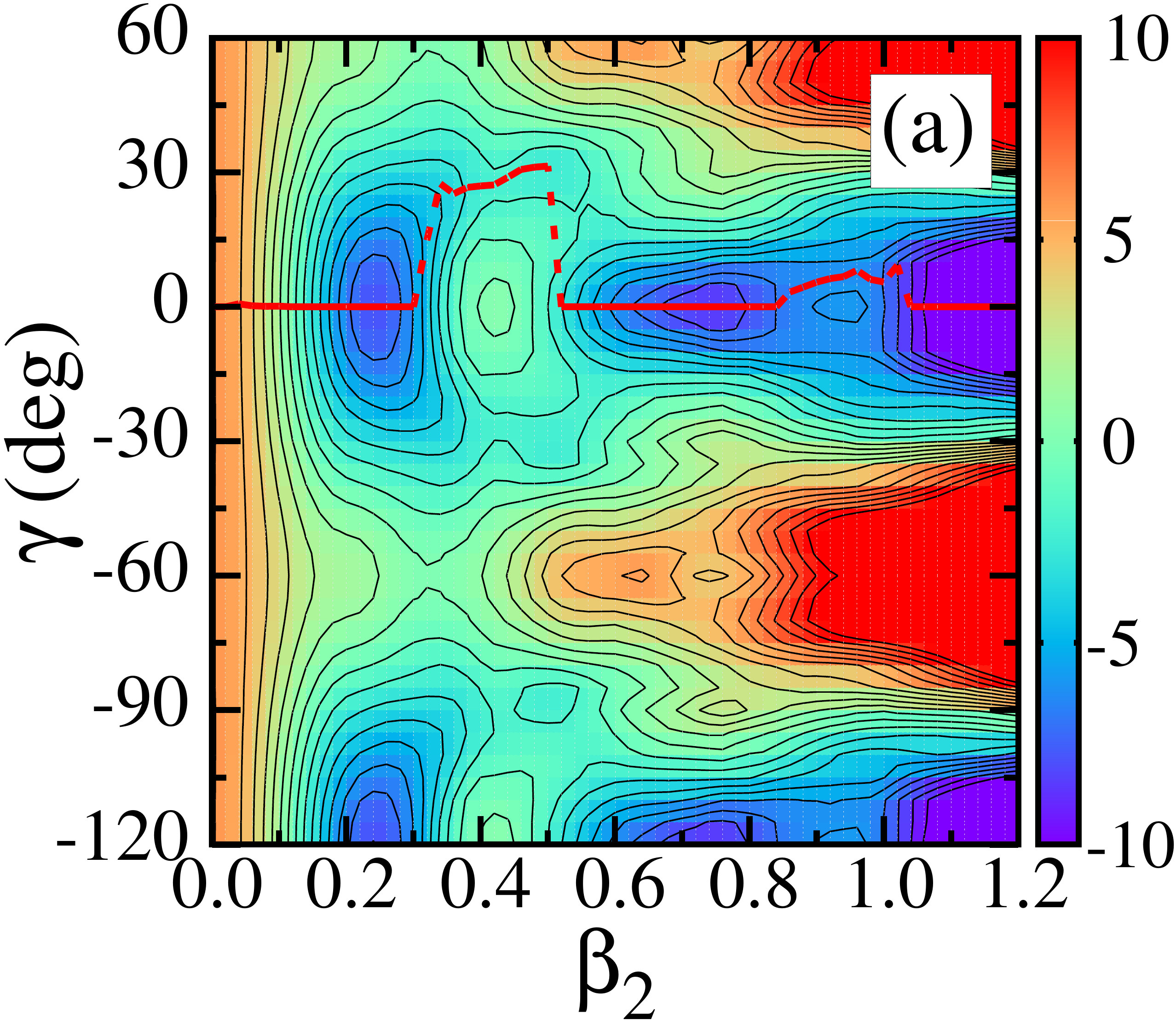}
\includegraphics[width=0.45\textwidth]{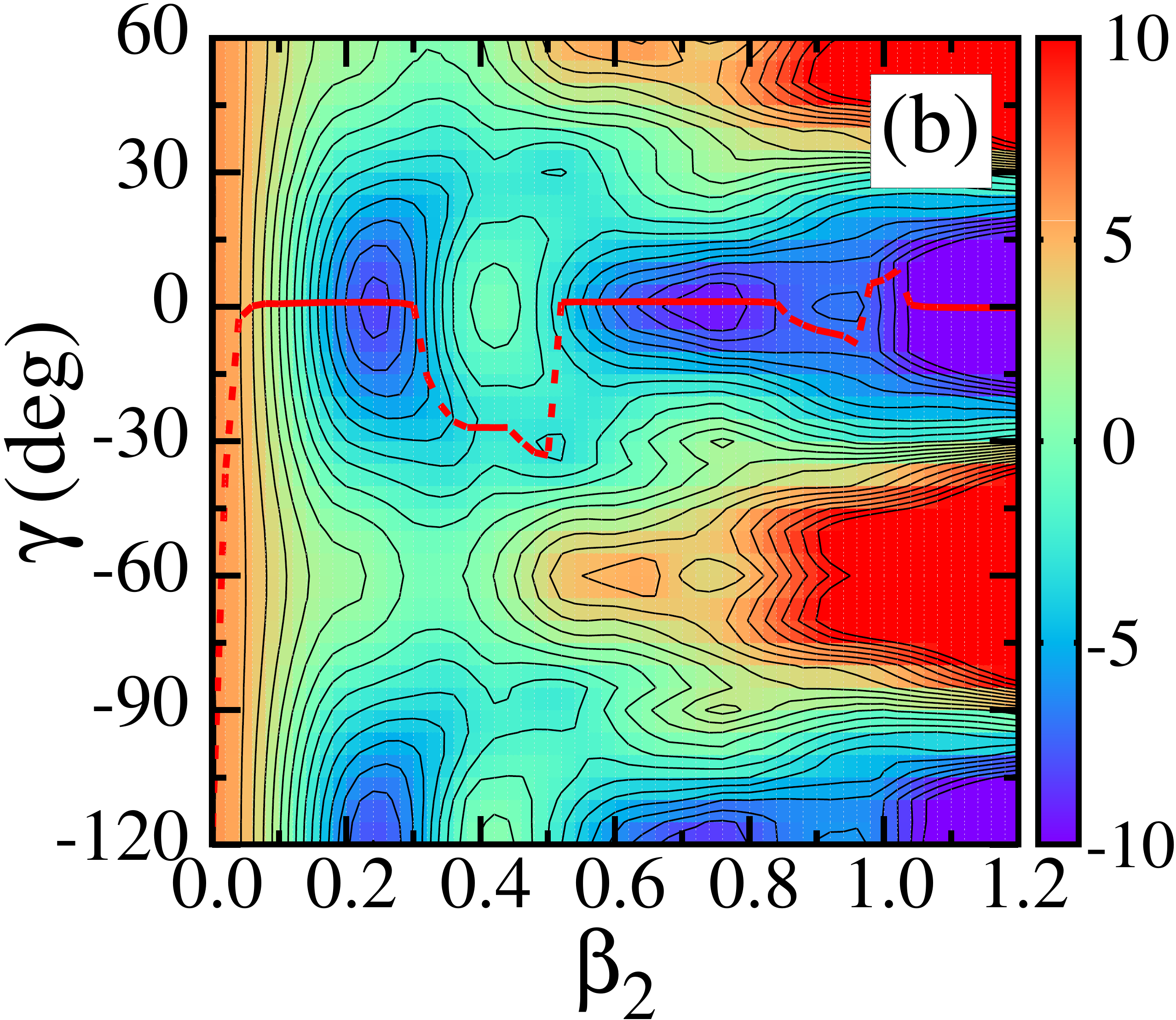}
\includegraphics[width=0.45\textwidth]{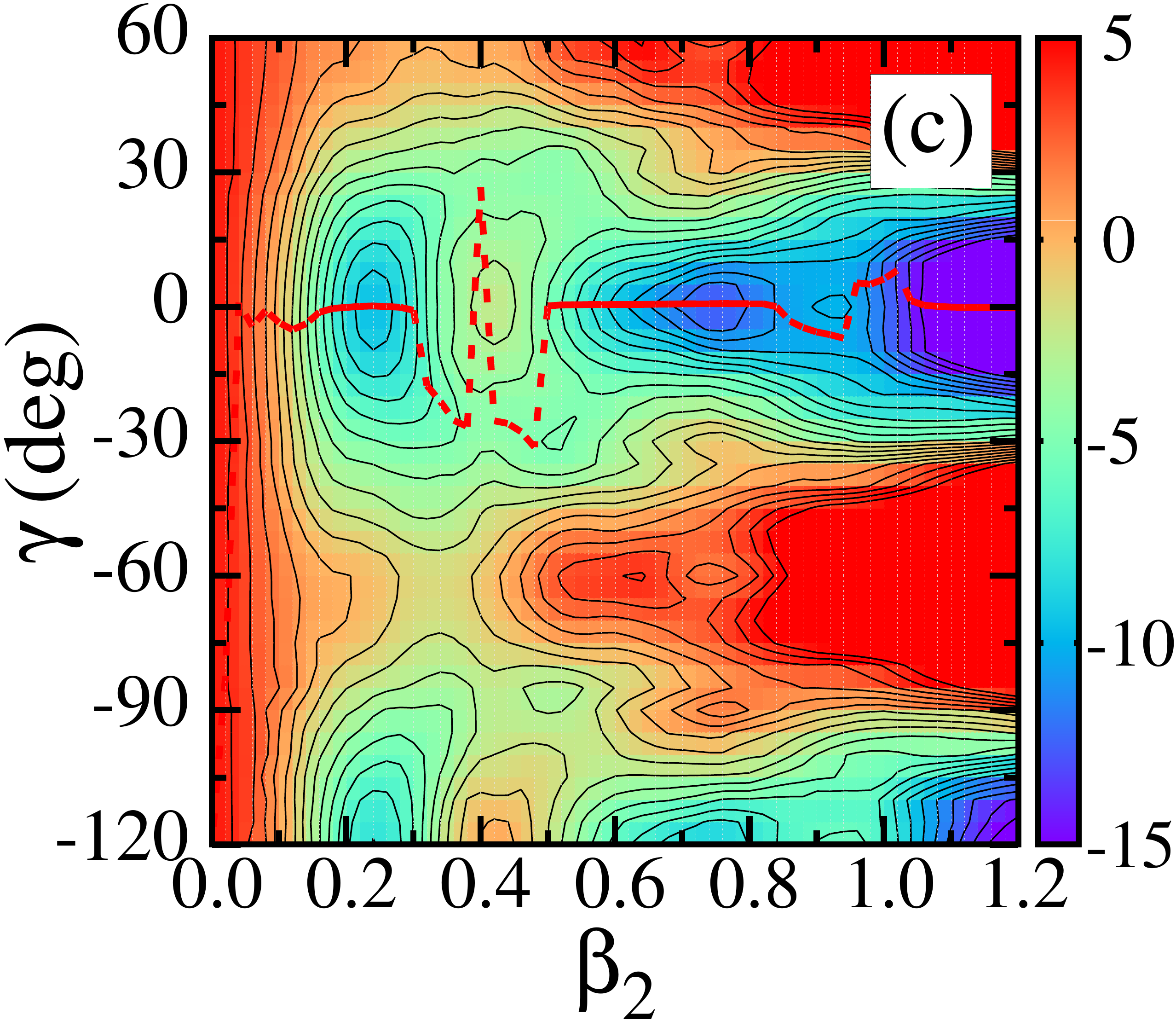}
\includegraphics[width=0.45\textwidth]{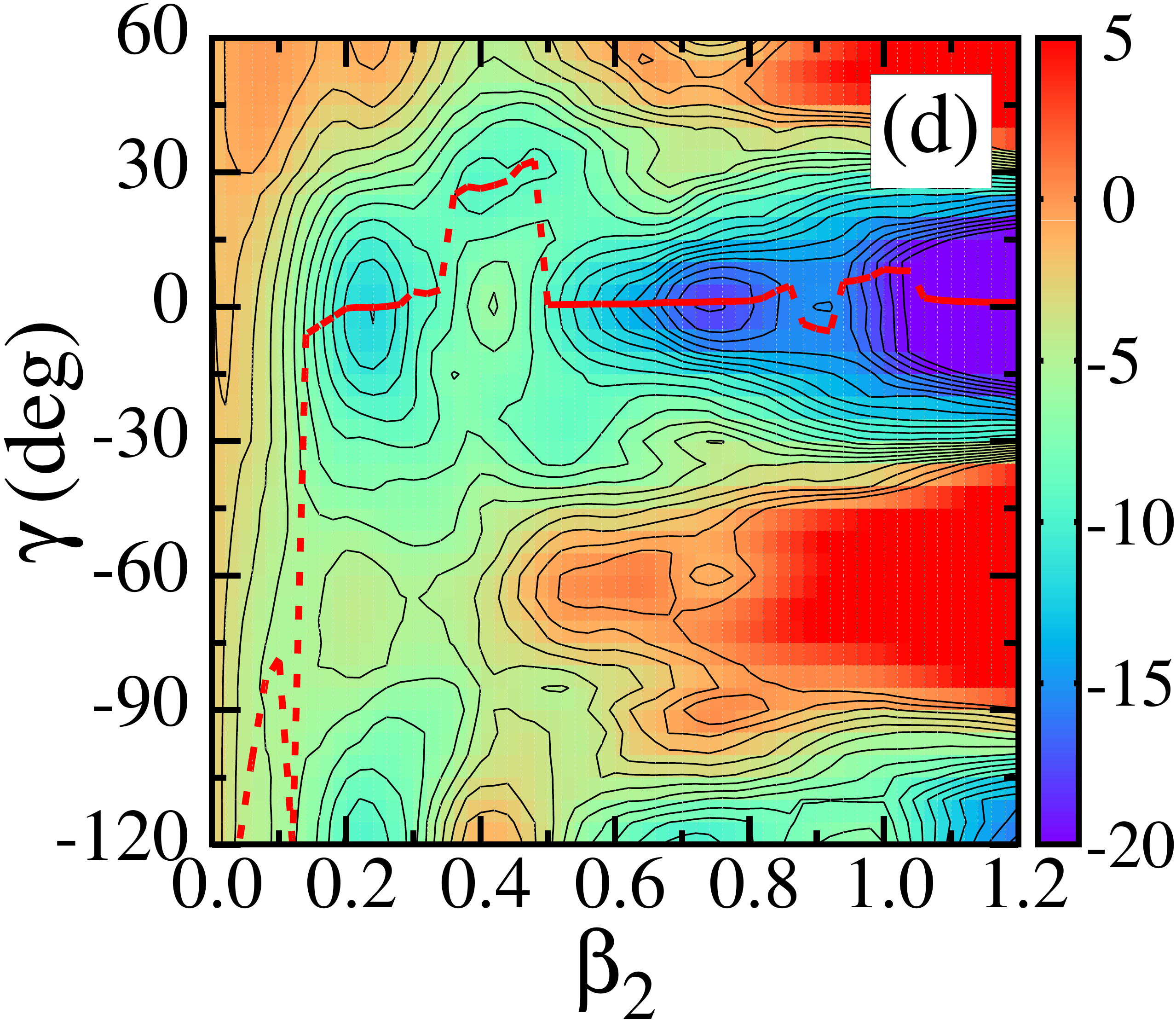}
\caption{Similar to Fig.~\ref{Fig1}(d) but for total Routhian projections of
$^{256}_{106}$Sg$_{150}$ at rotational frequencies $\hbar\omega=0.0$ (a),
0.1 (b), 0.2 (c) and 0.3 (d) MeV, respectively.}
                                                     \label{Fig8}
\end{figure*}

Figure~\ref{Fig8} illustrates the total Routhian surfaces projected
on the $\beta$ vs $\gamma$ plane for $^{256}_{106}$Sg$_{150}$ at
several typical rotational frequencies. At each grid in the maps,
the minimization of the total Routhian was performed over $\beta_4$.
It needs to be stressed that the energy domains denoted by the color
palettes are different in Figs.~\ref{Fig8}(c) and (d) for a better
display. Under rotation, the triaxial deformation parameter $\gamma$
covers the range from $-120^\circ$ to $60^\circ$ because the three
sectors ($-120^\circ$, $-60^\circ$), ($-60^\circ$, $0^\circ$) and
($0^\circ$, $60^\circ$) will represent rotation about the long,
medium and short axes, respectively (the nucleus with triaxial
shape). The nucleus with four $\gamma$ values $-120^\circ$,
$-60^\circ$, $0^\circ$ and $60^\circ$ has the axially symmetric
shape but different rotational orientation (cf. e.g.,
Ref.~\cite{Wang2015}). For instance, the triaxial deformation
parameter $\gamma = -120^\circ$ during rotation denotes that a
prolate nucleus with a non-collective rotation (namely, rotating
around its symmetry axis; see, e.g., the low-frequency part on the
fission path in Fig.~\ref{Fig8}(d)).  The 1D cranking is limited in
the present study. From this figure, one can see the evolution
properties of the triaxiality and rotation axis for both the
equilibrium shape and states along the fission path.
%
%
%
%
%
%

\begin{figure}[htbp]
\centering
\includegraphics[width=0.45\textwidth]{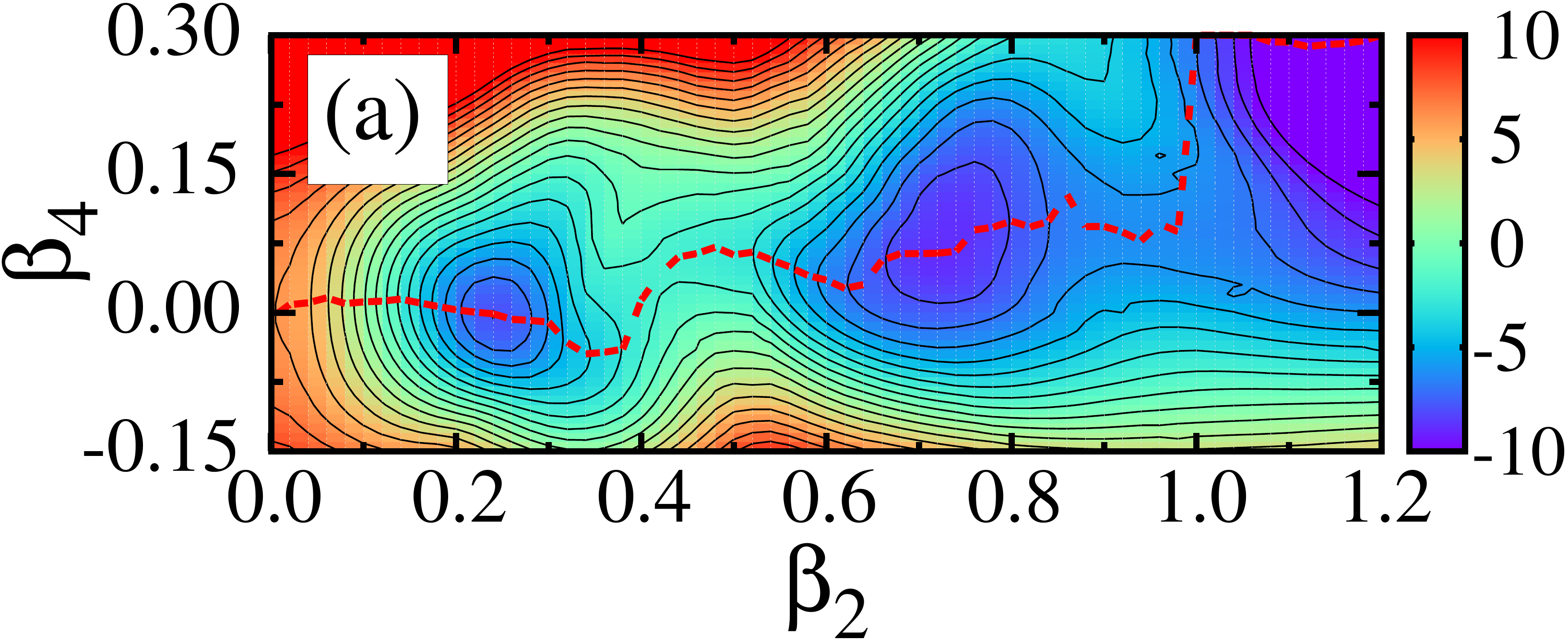}
\includegraphics[width=0.45\textwidth]{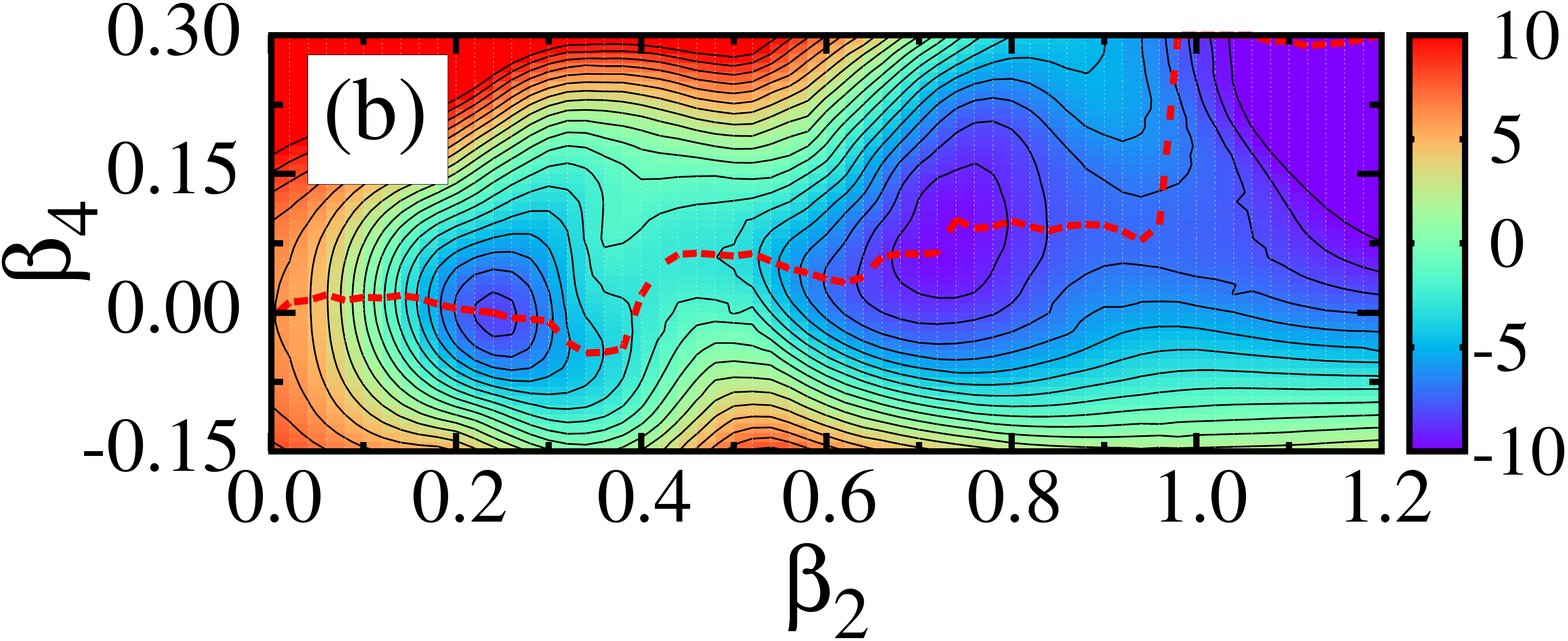}
\includegraphics[width=0.45\textwidth]{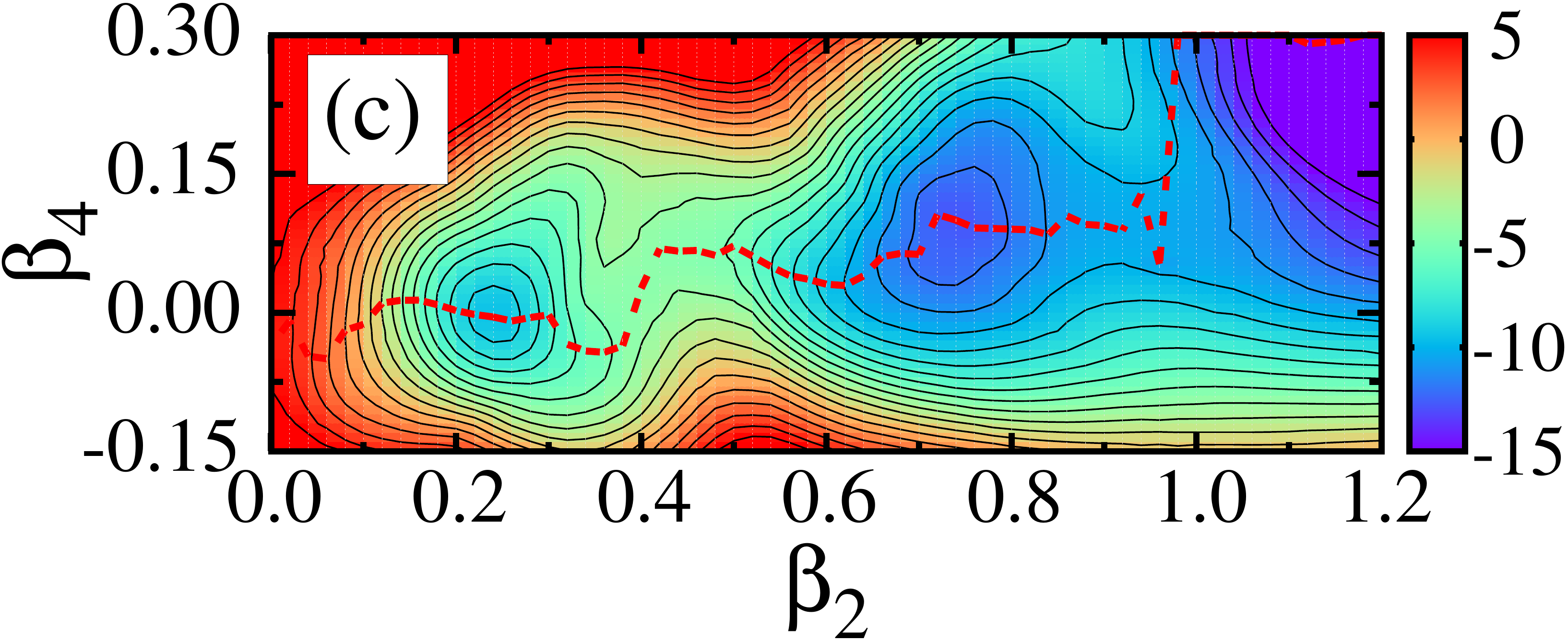}
\vspace{-0.5cm}
\includegraphics[width=0.45\textwidth]{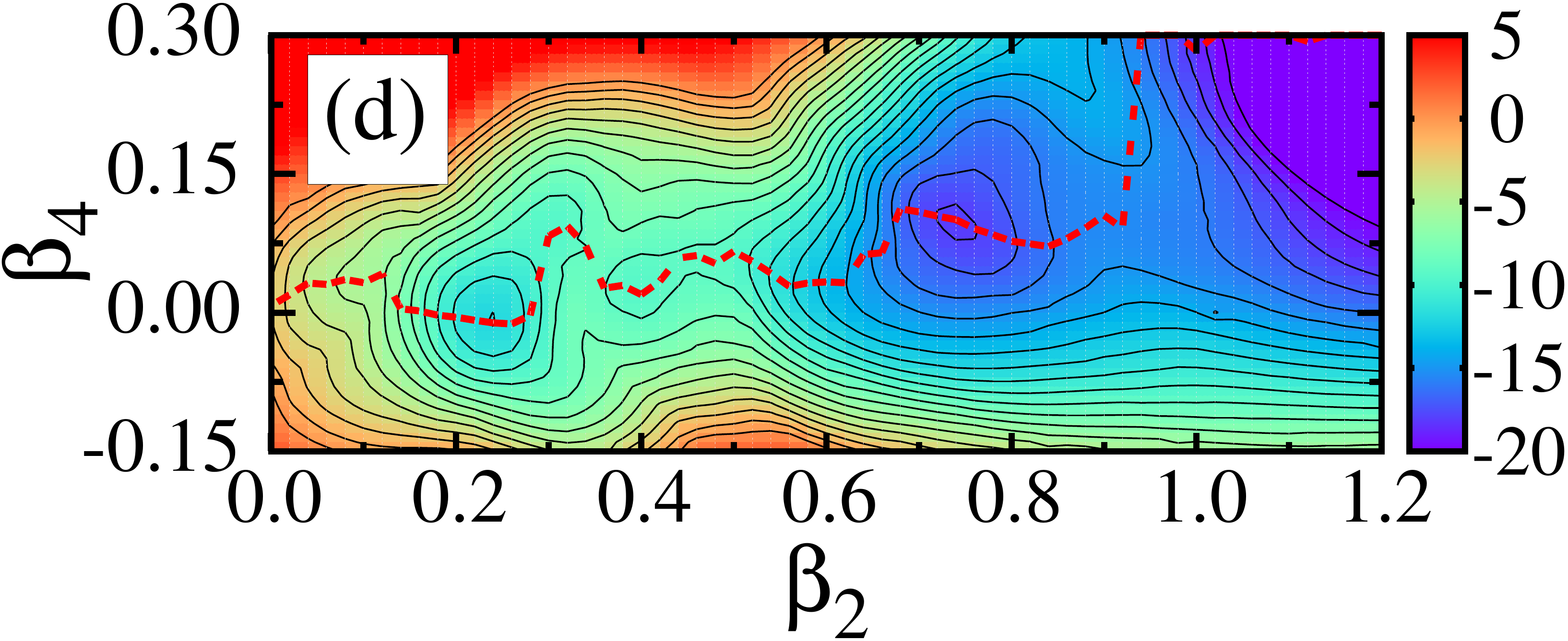}
\caption{Similar to Fig.~\ref{Fig2}(h) but for total Routhian projections of
$^{256}_{106}$Sg$_{150}$ at rotational frequencies $\hbar\omega=0.0$ (a),
0.1 (b), 0.2 (c) and 0.3 (d) MeV, respectively.}
                                                     \label{Fig9}
\end{figure}
%

%
%
%
%
%
%

\begin{figure}[htbp]
\centering
\includegraphics[width=0.48\textwidth]{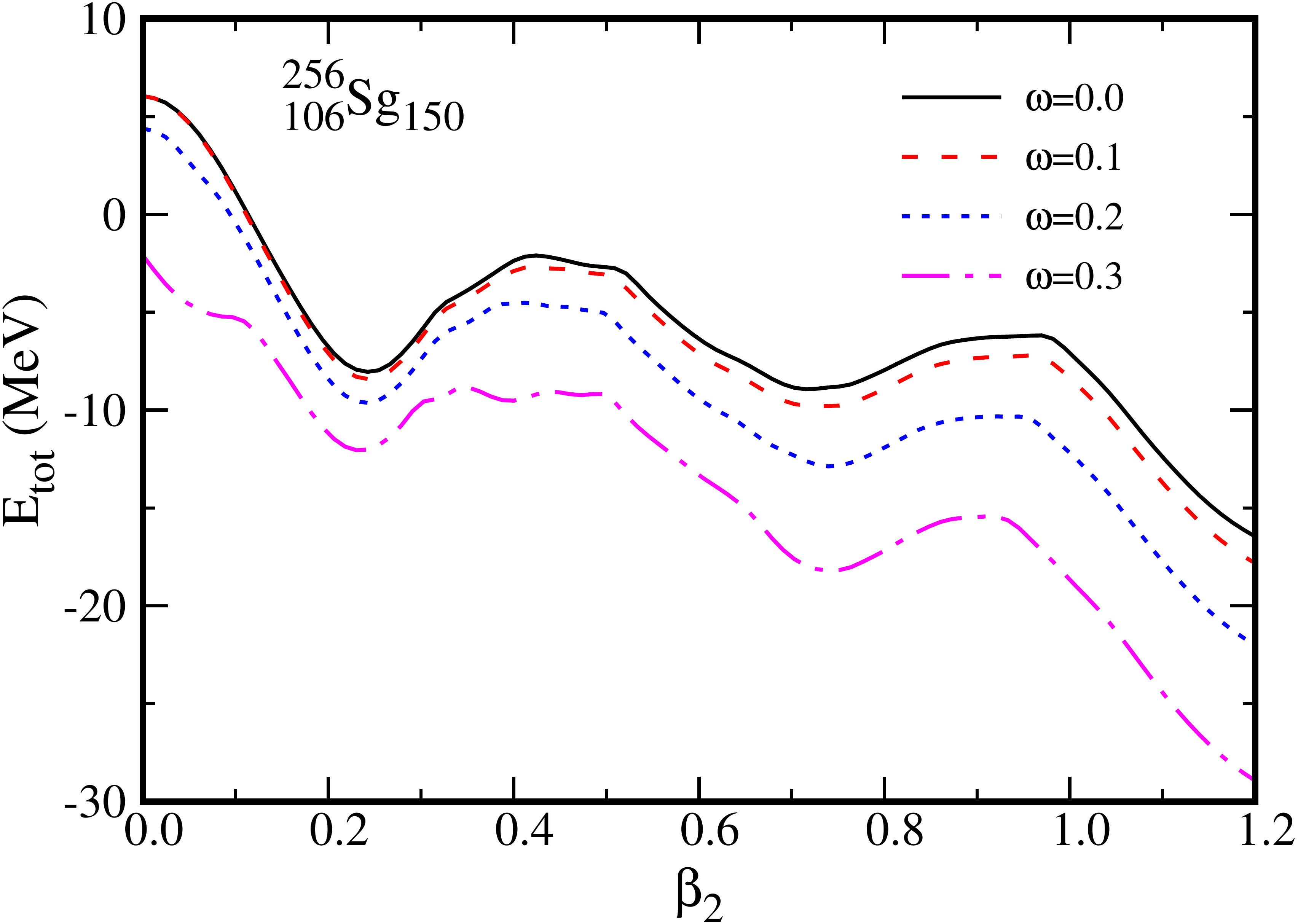}
\caption{The calculated total Routhian curves against $\beta_2$ for $^{256}_{106}$Sg$_{150}$ at four
selected frequencies $\hbar\omega=0.0$,
0.1, 0.2 and 0.3 MeV. At each $\beta_2$ point, the minimization was performed over $\gamma$ and $\beta_4$.}
                                                     \label{Fig10}
\end{figure}

To investigate the hexadecapole-deformation effect under rotation,
the total Routhian surfaces projected on the ($\beta_2$, $\beta_4$)
plane are shown in Fig.~\ref{Fig9} for $^{256}_{106}$Sg$_{150}$ at
four selected rotational frequencies $\hbar\omega=0.0$, 0.1, 0.2 and
0.3 MeV, respectively. Note that the color palletes are slightly
adjusted, similar to those in Fig.~\ref{Fig8}. It can be seen that
the hexadecapole deformation $\beta_4$ can strongly decrease the
total Routhian along the fission path, especially at high rotational
frequency and large quadrupole deformation. In other words, the
fission pathway will be modified by the the hexadecapole deformation
$\beta_4$. It should be pointed out that from this figure one can
find that part of the fission pathway evolutes along the border
(with $\beta_4$=0.30) of the calculation domain, indicating the
nucleus may possess a larger $\beta_4$ at this moment.
Figure~\ref{Fig10} illustrates the total Routhian curves in
functions of $\beta_2$ for $^{256}_{106}$Sg$_{150}$ at the selected
rotational frequencies mentioned above. The size and shape of the
inner and outer barriers and their evolution with rotation can be
evaluated conveniently. In the previous studies, e.g., in
Refs.~\cite{Wang2012,Lu2014,Kostryukov2021} , it was pointed out
that the octupole correlation may further decrease the outer barrier
in this mass region based on the PES calculation and fission
fragment analysis. The outer barrier for this nucleus may finally be
very low. It will be an open problem whether it will be able to play
a certain role in blocking the fission process.
%
%
%
%
%


\section*{4. Conclusions}
\label{conclusions}

We evaluate the structure evolution along the fission pathway for
$^{256}$Sg by using the multi-dimensional potential-energy(or
Routhian)-surface calculations, focusing on the effects of triaxial
and hexadecapole deformation and Coriolis force. Nuclear shape and
microscopic single-particle structure are investigated and analyzed.
The present results are compared with other theories. The properties
of nuclear shape and fission barrier are analyzed by comparing with
its neighboring even-even nuclei, showing a reasonable agreement.
Based on the deformation energy or Routhian curves, the fission
barriers are analyzed, focusing on their shapes, heights, and
evolution with rotation. It is found that the triaxial deformation
$\gamma$ decreases the potential energy on the landscape near the
saddles but the hexadecapole deformation $\beta_4$ (especially the
axial $\alpha_{40}$ component) modifies the least-energy fission
path after the first minimum, especially in $^{256}$Sg. In addition,
in contrast to the inner barrier, the outer barriers seem to have an
increasing trend from $^{260}$Sg to $^{256}$Sg which may be benefit
for blocking the fission of $^{256}$Sg to some extent. Next, it will
be necessary to simultaneously consider the reflection asymmetry in
a more reasonable deformation subspace.

\section*{Acknowledgement}

This work was supported by the National Natural Science Foundation
of China (Nos. 11975209, U2032211 and 12075287), the Physics
Research and Development Program of Zhengzhou University (No.
32410017), and the Project of Youth Backbone Teachers of Colleges
and Universities of Henan Province (No. 2017GGJS008). Some of the
calculations were conducted at the National Supercomputing Center in
Zhengzhou.

\section*{Conflict of Interest}

The authors declare that they have no known
competing financial interests or personal relationships that could have
appeared to influence the work reported in this paper.


\newpage


\end{document}